\documentclass[twoside]{article} 
 
\usepackage{amsmath,amsthm,amstext}
\usepackage{amssymb}

\input xy
\xyoption{all}

\long\def\comment#1{}

\newtheorem{theo}{Proposition}
\newtheorem{Theo}{Theorem}
\newtheorem{Coro}{Corollary}[Theo]
\newtheorem{remark}[Theo]{Remark}

\newtheorem{Dfn}{Definition} 
\newcounter{punkt}
\newenvironment{defn}[1]{\begin{Dfn}[#1]\normalfont\setcounter{punkt}{1}}%
{\setcounter{punkt}{1}\end{Dfn}}
\newcommand{\Bdef}[1]{\begin{defn}{#1}}
\newcommand{\Edef}{\end{defn}}

\theoremstyle{definition}

\newcommand{\NR}{\mathbb{R}}

\newcommand{\NC}{\mathbb{C}}

\newcommand{\CA}{\cal{A}}
\newcommand{\CH}{\cal{H}}
\newcommand{\CP}{\cal{P}}

\mathchardef\za="710B  %\alpha
\mathchardef\zb="710C  %\beta
\mathchardef\zg="710D  %\gamma
\mathchardef\zd="710E  %\delta
\mathchardef\ze="710F %\epsilon
\mathchardef\zz="7110  %\zeta
\mathchardef\zh="7111  %\eta
\mathchardef\zy="7112 %\theta
\mathchardef\zi="7113  %\iota
\mathchardef\zk="7114  %\kappa
\mathchardef\zl="7115  %\lambda
\mathchardef\zm="7116  %\mu
\mathchardef\zn="7117  %\nu
\mathchardef\zx="7118  %\xi
\mathchardef\zp="7119  %\pi
\mathchardef\zr="711A  %\rho
\mathchardef\zs="711B  %\sigma
\mathchardef\zt="711C  %\tau
\mathchardef\zu="711D  %\upsilon
\mathchardef\zf="711E %\phi
\mathchardef\zq="711F  %\chi
\mathchardef\zc="7120  %\psi
\mathchardef\zw="7121  %\omega
\mathchardef\zve="7122  %\varepsilon
\mathchardef\zvy="7123  %\vartheta
\mathchardef\zvo="7124  %\varomega
\mathchardef\zvr="7125 %\varrho
\mathchardef\zvs="7126 %\varsigma
\mathchardef\zvf="7127  %\varphi
\mathchardef\zG="7000  %\Gamma
\mathchardef\zD="7001  %\Delta
\mathchardef\zY="7002  %\Theta
\mathchardef\zL="7003  %\Lambda
\mathchardef\zX="7004  %\Xi
\mathchardef\zP="7005  %\Pi
\mathchardef\zS="7006  %\Sigma
\mathchardef\zU="7007  %\Upsilon
\mathchardef\zF="7008  %\Phi
\mathchardef\zC="7009  %\Psi
\mathchardef\zW="700A  %\Omega
\def\p{{\partial}}

\def\cpn{\NC{P^{N}}}
\def\cp#1{\NC{P^{#1}}}

\newcommand{\be}{\begin{equation}}
\newcommand{\ee}{\end{equation}}
\newcommand{\ba}{\begin{array}}
\newcommand{\ea}{\end{array}}
\newcommand{\fun}[3]{\mbox{${#1}: {#2}\rightarrow {#3}$}}
\newcommand{\su}[1]{\mathop{\mathfrak{su}({#1})}\nolimits}
\newcommand{\sltwo}{\ifmmode \ali{sl}{2} \else $\ali{sl}{2}$\fi}
\def\SU#1{\gr{SU}{#1}}
\newcommand{\Mat}{\textbf{M}(2,\,\NC)}

\newcommand{\hp}{\hphantom{-{}}}
\newcommand{\ol}[1]{\overline{#1}}

\newcommand{\BC}[1]{\NC^{#1}}

\def\ie{i.e.}
\def\eg{e.g.}

\newcommand{\fr}[1]{\mathfrak{#1}}

\newcommand{\gr}[2]{\mathop{\mathbf{#1}(#2)}\nolimits}
\newcommand{\gru}[3]{\mathop{\mathbf{#1}{(#2,\,#3)}}\nolimits}
\newcommand{\ali}[2]{\mathop{\fr{#1}(#2)}\nolimits}

\newcommand{\tr}{\mathop\mathrm{tr}\nolimits}
\newcommand{\id}{\mathop\mathrm{id}\nolimits}
\newcommand{\qmat}[1]{\left[\begin{matrix} #1 \end{matrix}\right]}

\newcommand{\IS}[2]{\sum\limits_{#1}^{#2}}

\newcommand{\DER}[3]{\dfrac{d^{#1}#2}{d{#3}^{#1}\hfill}}
\def\set#1#2{\{\,#1\mid#2\,\}}

\begin{document}

\title{ \vspace*{-3cm} $\mathbb{C}P^N$ sigma models via the $\SU{2}$ coherent states approach*}

\author{A. M. Grundland \\ 
Centre de Recherches Math\'ematiques, Universit\'e de Montr\'eal,\\
Montreal, C.P. 6128 (Qc) H3C 3J7, Canada\\
Department of Mathematics and Computer Sciences,\\
Universit\'e du Qu\'ebec, Trois-Rivi\`eres, CP 500 (QC) G9A 5H7, Canada\\
E-mail: grundlan@crm.umontreal.ca 
\and A. Strasburger \\
Institute of Mathematics and Cryptology, Faculty of
Cybernetics,\\
Military University of Technology --- WAT,\\
ul. S. Kaliskiego 2, 00-908 Warsaw, Poland\\
E-mail: aleksander.strasburger@wat.edu.pl 
\and
D. Dziewa-Dawidczyk \\
Department of Applied Mathematics,\\
Warsaw University of Life Sciences 
(SGGW)\\ 
ul.Nowoursynowska 159, 02-787 Warszawa, POLAND\\ 
E-mail: Diana\_Dziewa\_Dawidczyk@sggw.pl}

\maketitle  

\begin{small}
*To be published in Banach Center Proceedings\\
\end{small} 

\begin{abstract} 
%\noindent 

In this paper we present results obtained from the unification of
$\SU{2}$ coherent states with $\cpn$ sigma models defined
on the Riemann sphere having  finite actions. The set of coherent states
generated by a vector belonging to a carrier space of an irreducible
representation of the group gives rise to a map from the sphere into the
set of rank-1 Hermitian projectors in that space. The map can
be identified with a particular solution of the $\cpn$ sigma
model, where $N+1$ is equal to the dimension of the representation space.
In particular a choice of the generating vector as the highest weight vector of the
representation gives rise to the map known as a Veronese immersion. Using
a description of the matrix elements of these representations in terms of Jacobi polynomials, we obtain an explicit parametrization of the
solutions of the $\cpn$ models, which has not been previously
found. We relate the analytical properties of the solutions, which are
known to belong to separate classes --- holomorphic, anti-holomorphic and
various types of mixed ones ---  to the weight corresponding to the chosen
weight vector. Some examples of the described constructions are elaborated
in detail in this paper.

\end{abstract}

\section{Introduction}

In this paper we study properties of solutions of $\cpn$ sigma models in the group theoretical perspective, in particular using the language and methods of the theory of generalized coherent states. Such an analysis is facilitated by the introduction of a Hermitian projector formalism into the formulation of $\cpn$ sigma models, \cite{WZ},  and the recognition of the group theoretical character of some of the solutions of $\cpn$ sigma models in \cite{GU1}. However, the interpretation within the framework of the theory of coherent states seems to have been noticed only recently, \cite{AS}, adding a new application to this rapidly developing area of modern physics (see e.g. \cite{JPG} and references therein). 
 
We show that the systems of coherent states generated by the weight vectors of an irreducible representation of the $\SU{2}$  group can be described as a map into Hermitian rank-one projectors giving rise to harmonic transforms of the Veronese surface in the projective space associated to with the carrier space of the representation. Further, we formulate the invariant recurrence relations for $\cpn$ models (\cite{GG}) in terms of shift operators for the representation.  We also point to the possibility of parametrizing solutions of the $\cpn$ model in terms of Jacobi polynomials. The connection between these two analytical descriptions constitute the main goal of this paper. It allows us to provide a unification of the $\SU{2}$ coherent states with the surfaces associated with $\cpn$ sigma models, immersed in the $\su{N+1}$ algebra.

This paper is organized as follows. In section 2 we introduce basic notions on $\cpn$ sigma models where we focus on the use of the projector formalism and the generalized Weierstrass formula for the immersion of surfaces into $\su{N+1}$ Lie algebras. In section 3 we recall the main elements of the $\cpn$ theory from the group-theoretical point of view, which allows us to perform further computations, including the introduction of coherent states and covariant maps. Then, an explicit parametrization of coherent states for spin representations of $SU(2)$ are expressed in terms of the Jacobi polynomials. Section 4 contains final remarks and possible future developments.

\section{Solutions of $\cpn$ models expressed in terms of projectors} 

\label{Sec:prelim}
The dynamical fields in the $\cpn$  sigma models are maps from the unit sphere $S^2$ to the complex projective space  $\cpn$. Such maps can be described in terms of functions (fields) \fun{z=(z_0,\,z_1,\,\ldots,\,z_n)}{S^2}{S^{2N+1}} taking values in the unit sphere $S^{2N+1}=\{z\in \BC{N+1}\mid |z|^2=1\}$, where the norm  $|z|= \langle{ z^\dagger,\,  z}\rangle ^{1/2}$ is derived from the standard Hermitian inner product 
 $\langle{z,\,w}\rangle = z^\dagger\cdot w =\sum_{j=0}^{N}\overline{z}_jw_j$.
 
The independent variables of the fields of the $\cpn$ model are pairs $(\xi^1,\xi^2)\in\mathbb{R}^2$ often written in complex form by $\xi=\xi^1+i\xi^2$, and its complex conjugate $\ol{\xi} =\xi^1-i\xi^2$.
The covariant derivatives $D_\mu$ of the field $z \in S^{2N+1} $ are given as 
\begin{equation}
 D_\mu z=\partial_\mu z-(z^\dagger\cdot\partial_\mu z)z,\qquad \partial_\mu=\partial_{\xi^\mu},\quad \mu=1,2.
\end{equation}
%Here $z^\dagger$ is the Hermitian conjugate of $z$. 
The dynamics of the $\cpn$ sigma model defined on the Riemann sphere $S^2=\mathbb{C}\cup\lbrace\infty\rbrace$ are determined by the stationary points of the action functional  $\mathcal{A}=\ \int_{S^2}\mathcal{L}(z) d\xi d\ol{\xi}$,  where the Lagrangian density $\mathcal{L}$ is (see e.g. \cite{WZ})
\begin{equation}
 \mathcal{L}(z)=\frac{1}{4}(D_\mu z)^\dagger\cdot(D_\mu z).
\end{equation} 
The Euler--Lagrange (E-L) equations take the form
\begin{gather}
 D_\mu D_\mu z+(D_\mu z)^\dagger\cdot(D_\mu z)z=0,.\label{eq5}
\intertext{subject to the algebraic constraint} 
z^\dagger z=1.
\end{gather}
Since $\mathcal{L}(z)$ is constant on fibers of the Hopf fibration $S^1\rightarrow S^{2N+1}\rightarrow\mathbb{C}P^N$, obtained form the circle $S^1$ action on $S^{2N+1}$ by coordinatewise multiplication, we observe that $\mathcal{L}(z)$ depends only on the map $[z]:\Omega\subset\mathbb{C}\rightarrow\mathbb{C}P^N$, where for $z\in S^{2N+1}$ we denote by
$[z]=\lbrace e^{i\psi}z\vert\psi\in\mathbb{R}\rbrace$ 
the element of the projective space $\cpn$ corresponding to $z$.

However, it is convenient to describe the models \eqref{eq5} in terms of the homogeneous, ``unnormalized'' field $\xi\mapsto f=(f_0,...,f_N)\in\mathbb{C}^{N+1}\backslash\lbrace0\rbrace$ related to the ``$z$'s'' for which
\begin{equation}
 z=\frac{f}{(f^\dagger\cdot f)^{1/2}},
\end{equation}
Using the standard notion of complex derivatives $\partial$ and $\ol{\partial}$  with respect to $\xi$ and $\ol{\xi}$ given by    
\begin{equation*}
 \partial=\frac{1}{2}\left(\frac{\partial}{\partial\xi^1}-i\frac{\partial}{\partial\xi^2}\right),\qquad \ol{\partial}=\frac{1}{2}\left(\frac{\partial}{\partial\xi^1}+i\frac{\partial}{\partial\xi^2}\right), 
\end{equation*}
we see that the homogeneous variable $f$ satisfies an unconstrained form of the E--L equations
\begin{equation}
 \left(\mathbb{I}_{N+1}-\frac{f\otimes f^\dagger}{f^\dagger\cdot f}\right)\cdot\left[\partial\bar{\partial}f-\frac{1}{f^\dagger\cdot f}\left((f^\dagger\cdot\bar{\partial}f)\partial f+(f^\dagger\cdot\partial f)\bar{\partial}f\right)\right]=0,\label{eq7}
\end{equation}
where $\mathbb{I}_{N+1}$ is the $(N+1)\times (N+1)$ identity matrix. 

An even more compact form of the E--L equations \eqref{eq7} is obtained by expressing them as a conservation law for the rank-one Hermitian projector% $P:S^2\rightarrow\mbox{Aut}(\mathbb{C}^N)$
\begin{equation}
 P=\frac{f\otimes f^\dagger}{f^\dagger\cdot f},\label{eq8}
\end{equation}
satisfying $P^2=P$, $P^\dagger=P$ associated with the field $f$.
In this formulation the action functional takes the form 
%The solutions of the $\cpn$ model are stationary points of the 
\begin{equation}
 \mathcal{A}(P)=\int_{S^2}\tr(\partial P\ol{\partial}P)d\xi d\bar{\xi}.\label{action}
\end{equation}
 
In terms of the projector $P$ the E--L equations \eqref{eq5} take the simple form 
\begin{equation}
 [\partial\bar{\partial}P,P]=0,\label{eq10}
\end{equation}
or equivalently can be written as the conservation law
\begin{equation}
 \partial [\bar{\partial}P,P]+\bar{\partial}[\partial P,P]=0.\label{eq9_CL}
\end{equation}
This gives the following expression for the matrix-valued (or more precisely $\su{N+1}$-valued) $1$-form 
\begin{equation}
dX= i(-[\ol{\partial}P,P] d\xi + [\partial P,P]d\ol{\xi} ) \label{eq9_form}
\end{equation}
which is closed and therefore can be integrated in any simply connected domain, \eg~ on the surface of the sphere $S^2$, leading to an immersion of the domain into the Lie algebra $\su{N+1}$. We will expand on that point later on.  

Based on the Gram-Schmidt orthogonalization procedure, a method for constructing an entire class of solutions admitting the finite action $\CA$  of the $\mathbb{C}P^{N}$ model  was proposed by A. Din and W.J. Zakrzewski \cite{DZ},  later studied  by R. Sasaki \cite{Sasaki}, and improved by  Eells and Wood \cite{EW}. 

Under the assumption that the model is defined on the Riemann sphere $S^2=\mathbb{C}\cup\lbrace\infty\rbrace$ and that its action \eqref{action} is finite, all the solutions can be obtained from a basic  
solution given in terms of the holomorphic nonconstant function by successive applications of the ``raising'' operator:
\begin{equation}
 P_+:f\in\mathbb{C}^{N+1}\backslash\lbrace0\rbrace\rightarrow P_+f=\left(\mathbb{I}_{N+1}-\frac{f\otimes f^\dagger}{f^\dagger\cdot f}\right)\partial f,\qquad \text{for}\quad\ol{\partial}f=0\label{eq10a}
\end{equation}
or analogously in terms of a basic antiholomorphic solution given by  an antiholomorphic nonconstant function under the application of  the ``lowering'' operator:
\begin{equation}
 P_-:f\in\mathbb{C}^{N+1}\backslash\lbrace0\rbrace\rightarrow P_-f=\left(\mathbb{I}_{N+1}-\frac{f\otimes f^\dagger}{f^\dagger\cdot f}\right)\ol{\partial} f,\qquad \text{for}\quad \partial f=0. \label{eq10b}
\end{equation}
 
This method allows us to construct three classes of solutions: holomorphic, anti-holomorphic and mixed solutions, which are determined by
\begin{equation}
 f_k:=P^k_+f,\quad k=0,1,...,N,\quad\mbox{where }P^0_+= \id,\quad P_+^{N+1}f=0.\label{eq11}
\end{equation}
Here the operator $P_+^k$ is obtained by applying the operator $P_+$ $k$ times successively. As a result we not only have information about all harmonic maps $S^2\mapsto\cpn$ but also an orthogonal basis in $\mathbb{C}^{N+1}$ of solutions of the $\cpn $ model \eqref{eq9_CL} \cite{Bob}.
 
The raising and lowering operators $P_\pm$ of solution \eqref{eq9_CL} can also be expressed in terms of projector operators through the formulas given in \cite{GG}.
\begin{align}
 %\begin{array}{l}
 & P_{k+1}=\Pi_+(P_k)=\dfrac{\ol{\partial}P_kP_k\partial P_k}{\mbox{tr}(\ol{\partial}P_kP_k\partial P_k)},\qquad k=0,1,...,N\label{recurrence1}\\%[6pt]
&  P_{k-1}=\Pi_-(P_k)=\dfrac{\partial P_kP_k\ol{\partial}P_k}{\mbox{tr}(\partial P_kP_k\ol{\partial}P_k)},\qquad \mbox{where }P_k=\dfrac{f_k\otimes f_k^\dagger}{f_k^\dagger\cdot f_k}.\label{recurrence2}\\
&\sum_{j=0}^NP_j=\mathbb{I}_N,\qquad P_kP_j=\delta_{kj}P_j.\label{orthocond}
\end{align}
As a result equation \eqref{eq8} gives an isomorphism between the equivalence classes of the $\cpn$ model and the set of rank-one Hermitian projectors $P_k$. To each of these solutions \eqref{eq11} we can associate a surface in the $\su{N+1}$ algebra \cite{GZ2} and, using equation \eqref{eq9_form} as in \cite{GY},  
we obtain a sequence of surfaces
\begin{equation}
 X_k=-i\left(P_k+2\sum_{j=0}^{k-1}P_j\right)+ic_k\mathbb{I}_{N+1}\in\mathfrak{su}(N+1), \quad c_k=\frac{1+2k}{N+1}.\label{surfacesk}
\end{equation}
Here, the $c_k$'s are integration constants, ensuring that the integrals  are skew-Hermitian and traceless, and therefore belong to the Lie algebra $\su{N+1}$.   The $\su{N+1}$-valued matrix functions $X_k(\xi,\ol{\xi})$ constitute the generalized Weierstrass formula for the immersion of 2D surfaces into $\NR^{N(N+2)}$, isomorphic to the Lie algebra $\su{N+1}$, \cite{Kono}.
The matrix-valued functions $X_k$ satisfy the following cubic matrix equations (the minimal polynomial identity) \cite{GG2},
\begin{equation}
 (X_k-ic_k\mathbb{I}_{N+1})(X_k-i(c_k-1)\mathbb{I}_{N+1})(X_k-i(c_k-2)\mathbb{I}_{N+1})=0,\qquad 0<k<N
\end{equation}
for any mixed solution of equation \eqref{eq9_CL} in the $\cpn$ model. For any holomorphic ($k=0$) or anti-holomorphic ($k=N$) solutions of equation \eqref{eq9_CL} in the $\cpn$ model, the minimal polynomial for the functions $X_0$ and $X_{N}$ is quadratic, 
\begin{align*} 
  (X_0-ic_0\mathbb{I}_{N+1})(X_0-i(c_0-1)\mathbb{I}_{N+1}) &=0,\\
  (X_{N}+ic_0\mathbb{I}_{N+1})(X_{N}+i(c_0-1)\mathbb{I}_{N+1})&=0,\qquad\mbox{where }c_0+c_{N}=2.
\end{align*} 
The projectors $P_k$ fulfill the completeness relation  
 $\sum_{k=0}^{N}P_k=\mathbb{I}_{N+1}$,
which implies in turn that the immersion functions $X_k$ satisfy the linear relation 
 $\sum_{k=0}^{N}(-1)^kX_k=0$.
The raising and lowering operators $\chi_\pm$ for the sequence of surfaces were devised in \cite{GG}. These operators map between the surfaces as follows
\begin{align}
&  X_{k+1}=\chi_+(X_k)=X_k-i(\Pi_+(P_k)+P_k)+\frac{2i}{N+1}\mathbb{I}_{N+1},\\
&  X_{k-1}=\chi_-(X_k)=X_k+i(\Pi_-(P_k)+P_k)-\frac{2i}{N+1}\mathbb{I}_{N+1}.
\end{align}
As a result a certain geometric characterization of the surfaces immersed in the $\su{N+1}$ algebra can be performed (see e.g. \cite{GSZ,GG,PG}). The geometric properties are such that the surfaces $X_k$ are conformally parametrized, and we can derive the 1st and 2nd fundamental forms, principal curvatures, topological charges, Willmore functionals  and Euler--Poincar\'e characteristics of the surfaces.

We would like to illustrate the above procedure by describing a particular solution of the  $\cpn$ sigma model, which will be discussed in detail in Section $3$, coming from the classical Veronese imbeddings, cf. \cite{Boal}. For the case of the $\mathbb{C}P^2$ model from the holomorphic Veronese map $f_0=(1,\sqrt{2}\xi,\xi^2)\in\NC^{3}\setminus \{0\}$ we obtain a sequence of projectors 
 \arraycolsep 2pt
\begin{align}
P_0&=\frac{f_0\otimes f_0^\dagger}{f_0^\dagger\cdot f_0}=\dfrac{1}{(1+\vert\xi\vert^2)^2}
\qmat{1 & 2^{1/2}\ol{\xi} & \ol{\xi}^2 \\
2^{1/2}\xi & 2\vert\xi\vert^2 & 2^{1/2}\vert\xi\vert^2\ol{\xi} \\
\xi^2 & 2^{1/2}\vert\xi\vert^2\xi & \vert\xi\vert^4}, \nonumber\\
P_1&=\frac{f_1\otimes f_1^\dagger}{f_1^\dagger\cdot f_1}=\dfrac{1}{(1+\vert\xi\vert^2)^2}
\qmat{2\vert\xi\vert^2 & 2^{1/2}(\vert\xi\vert^2-1)\ol{\xi} & -2\ol{\xi}^2 \\
2^{1/2}(\vert\xi\vert^2-1)\xi & (\vert\xi\vert^2-1)^2 & -2^{1/2}(\vert\xi\vert^2-1)\ol{\xi} \\
-2\xi^2 & -2^{1/2}(\vert\xi\vert^2-1)\xi & 2\vert\xi\vert^2},\label{projectors1} \\
P_2&=\frac{f_2\otimes f_2^\dagger}{f_2^\dagger\cdot f_2}=\dfrac{1}{(1+\vert\xi\vert^2)^2}
\qmat{\vert\xi\vert^4 & -2^{1/2}\vert\xi\vert^2\ol{\xi} & \ol{\xi}^2 \\
-2^{1/2}\vert\xi\vert^2\xi & 2\vert\xi\vert^2 & -2^{1/2}\ol{\xi} \\
\xi^2 & -2^{1/2}\xi & 1}.\nonumber 
%\end{split}
\end{align}
\arraycolsep 5pt
The corresponding $\su{3}$-valued forms give the following immersions for the surfaces  
\arraycolsep 2pt
\begin{align}
X_0 &=\dfrac{i}{(1+\vert\xi\vert^2)^{2}}
\qmat{\frac{1}{3}(\vert\xi\vert^4+2\vert\xi\vert^2-2) & -\sqrt{2}\ol{\xi} & -\ol{\xi}^2 \\
-\sqrt{2}\xi & \frac{1}{3}(\vert\xi\vert^4-4\vert\xi\vert^2+1) & -\sqrt{2}\vert\xi\vert^2\ol{\xi} \\
-\xi^2 & -\sqrt{2}\vert\xi\vert^2\xi & -\frac{1}{3}(2\vert\xi\vert^4-2\vert\xi\vert^2-1)},\nonumber\\
X_1 &=\dfrac{i}{(1+\vert\xi\vert^2)^2}
\qmat{\vert\xi\vert^2-1 & -\sqrt{2}\ol{\xi} & 0 \\
-\sqrt{2}\xi & 0 & -\sqrt{2}\ol{\xi} \\
0 & -\sqrt{2}\xi & -\vert\xi\vert^2+1)}, \label{immersions1} \\
X_2 &=\dfrac{i}{(1+\vert\xi\vert^2)^2}
\qmat{-\frac{1}{3}(1-2\vert\xi\vert^4+2\vert\xi\vert^2) & -\sqrt{2}\vert\xi\vert^2\ol{\xi} & \ol{\xi}^2 \\
-\sqrt{2}\vert\xi\vert^2\xi & -\frac{1}{3}(1+\vert\xi\vert^4-4\vert\xi\vert^2) & -\sqrt{2}\ol{\xi} \\
\xi^2 & -\sqrt{2}\xi & -\frac{1}{3}(\vert\xi\vert^4+2\vert\xi\vert^2-2)}.\nonumber
 \end{align}
\arraycolsep 5pt
 
Thus we see that the  only solutions of \eqref{eq9_CL} with a finite action \eqref{action}  are  given by holomorphic, mixed and antiholomorphic projectors $P_k$, $k=0,1,2$ obtained from the successive action of the contracting operator $P_+$ applied to $P_0$.

An advantage of the presented approach is that, without referring to any additional considerations, the recurrence relations give a very useful tool for constructing the sequence of successive surfaces $X_k$ associated with the $\cpn$ sigma model obtained from the knowledge of the previous one. The geometrical setting allows us to study certain global properties of successive surfaces $X_k$ as illustrated by the example of surfaces associated with the $\mathbb{C}P^2$ model.

We now demonstrate according to \cite{GG} that the surfaces $X_k$ and $X_l$ associated with the $\mathbb{C}P^{N-1}$ model do not intersect if $k\neq l$, with the exception of $X_0$ and $X_1$ in the $\mathbb{C}P^1$ model since $X_0$ and $X_1$ are equal in that case.

Proof. If $l$ and $k$ are two different indices of the induced surfaces, where $l>k$, then we obtain by subtracting \eqref{surfacesk} from the analogous expression for $X_l$
\begin{equation}
P_l-P_k+2\sum_{j=k}^{l-1}P_j-\frac{2(l-k)}{N}\mathbb{I}_N=0.\label{a1}
\end{equation}
Multiplying equation (\ref{a1}), by $P_k$ and using the orthogonality condition (\ref{orthocond}) we get
\begin{equation}
P_k\left(\mathbb{I}_N-\frac{2(l-k)}{N}\right)=0\label{a2}
\end{equation}
On the other hand, multiplying both sides of the expression \eqref{a1} by $P_{l-1}$ we obtain
\begin{equation}
P_{l-1}\left(\mathbb{I}_N-\frac{l-k}{N}\right)=0\qquad \mbox{for }k<l-1\label{a3}
\end{equation}
and
\begin{equation}
P_k\left(\mathbb{I}_N-\frac{2}{N}\right)=0\qquad \mbox{for }k=l-1\label{a4}
\end{equation}
The equations \eqref{a2}, \eqref{a3} and \eqref{a4} can only be satisfied when $N=2$, $l=1$ and $k=0$. This implies that $X_1=X_0$.\hfill$\square$

We now demonstrate that the immersion functions $X_k$ and $X_m$ make a constant angle according to the Euclidean inner product $(A,B)$ of the $\su{N}$ matrices
\begin{equation}
(A,B)=\frac{-1}{2}\mbox{tr}(A\cdot B).
\end{equation}
That is, the angle $\Phi_{km}$ between the immersion functions $X_k$ and $X_m$ is independent of the choice of projector $P_0$ and of the coordinates $\xi$ and $\bar{\xi}$. Hence, the angle $\Phi_{km}$ between two different immersion functions, $X_k$ and $X_m$ for $k<m$ is given by \cite{GG}
\begin{equation}
\cos\Phi_{km}=\frac{c_k(2-c_m)}{\lbrace [c_k(2-c_k)-1/N][c_m(2-c_m)-1/N]\rbrace^{1/2}}\label{aangle}
\end{equation}
The formula (\ref{aangle}) can be obtained either by using the scalar product from the generalized Weierstrass formula for immersion for $X_k$ and $X_m$ (\ref{surfacesk}) (taking into account the fact that the projectors $P_0,P_1,...,P_k$ are mutually orthogonal or by operating directly on the eigenvalues of the immersion functions). In both cases we obtain for $m>k$
\begin{equation}
(X_k,X_m)=\frac{-1}{2}\mbox{tr}(X_k\cdot X_m)=\frac{N}{2}c_k(2-c_m),
\end{equation}
or in the case when $k=m$
\begin{equation}
(X_k,X_k)=\frac{-1}{2}\mbox{tr}(X_k\cdot X_k)=\frac{1}{2}[Nc_k(2-c_k)-1]
\end{equation}
It is easy to show that $\cos\Phi_{km}\in(0,1)$ except in the case where $N=2$ (and clearly for $k=0$, $m=1$), for which the surfaces coincide and $\cos\Phi_{km}=1$. Equation (\ref{aangle}) is symmetric with respect to a transformation $k\leftrightarrow N-1-m$, as can be seen e.g. in tables of $\cos\Phi_{km}$ for the $\mathbb{C}P^2$ and $\mathbb{C}P^3$ models in sections 3.4 and 4.

Finally we wish to point out the relation of the above construction to the linear spectral problem (LSP) associated with the $\cpn$ sigma model as formulated in the papers \cite{ZM,DHZ} 
\begin{equation}
 \partial\Phi_k=\frac{2}{1+\lambda}[\partial P_k,P_k]\Phi_k,\qquad \ol{\partial}\Phi_k=\frac{2}{1-\lambda}[\ol{\partial}P_k,P_k]\Phi_k,\quad 0\leq k\leq N,\quad\lambda\in\mathbb{C}.\label{eq20}
\end{equation}
The explicit solutions $\Phi_k$ of the LSP (\ref{eq20}) which tend to identity matrix $\mathbb{I}$ as $\lambda\rightarrow\infty$, were found in \cite{DHZ} to be 
\begin{align}
&  \Phi_k=\mathbb{I}_{N+1}+\frac{4\lambda}{(1-\lambda)^2}\sum_{j=0}^{k-1}P_j-\frac{2}{1-\lambda}P_k\in SU(N+1),\qquad\lambda=it,\\
&  \Phi_k^\dagger=\Phi^{-1}_k=\mathbb{I}_{N+1}-\frac{4\lambda}{(1+\lambda)^2}\sum_{j=0}^{k-1}P_j-\frac{2}{1+\lambda}P_k,\qquad t\in\mathbb{R}.
\end{align}
The immersion functions  $X_k$  may be expressed in terms of the wavefunctions $\Phi_k$  in two ways, either by  the Sym--Tafel  immersion formula \cite{Sym} for completely integrable models
\begin{equation}
 X_k=\alpha(\lambda)\Phi^{-1}_k\partial_\lambda\Phi_k+ic_k\mathbb{I}_{N+1}\in\su{N+1},
\end{equation}
(where $\alpha(\lambda)$ is an arbitrary function of the spectral parameter $\lambda$), or using  its asymptotics for large values of $\lambda$ \cite{GG}, 
\begin{equation}
 X_k=\frac{i}{2}\lim_{\lambda\rightarrow\infty}[\lambda(\mathbb{I}_{N+1}-\Phi_k)]+ic_k\mathbb{I}_{N+1}\in\su{N+1}.
\end{equation}
So in conformal coordinates we obtain, as a result, a sequence of surfaces $X_k$ whose structural equations are identical to the equations of motion (\ref{eq9_CL}) for the $\cpn$ model, see \cite{GGP}
\begin{equation}
 [\partial\ol{\partial}X_k,X_k]=0,\qquad k=0,...,N.
\end{equation}

The main goal of this paper is to provide a unification of the $SU(2)$ coherent states and the surfaces associated with $\cpn$ sigma models immersed in the $\su{N+1}$ algebra.  Through this link, we derive the generating vectors as weight vectors of the representations in terms of the Jacobi polynomials parametrizing the solutions of the $\cpn$ model.

\section{Covariant construction of maps from $S^2$ into $\cpn$}

In order to establish the notation used in the sequel we shall recall in some detail the classical construction of the $\SU{2}$-covariant coverings of $S^2\simeq \NC{P^1}$ by the unit sphere $S^3\simeq \SU{2}$. Let 
\comment{The space of square $2\times 2$ matrices with complex entries will be denoted by $\Mat$. $\gru{SL}{2}{C}$ will stand for the group of unimodular $2\times 2$ matrices  with complex entries,}   
\be\label{eqn:sl_2}
\gru{SL}{2}{C}=\set{\qmat{a&b\\c&d}}{ a,b,c,d \in \NC,\ ad-bc= 1},  
\ee
and let $\SU{2}$ be its unitary  subgroup  
\be\label{eqn:su_2}
\SU{2}=\set{\qmat{u &-\ol{v}\\ v &\hp \ol{u}}}{u,v\in \NC,\ |u|^2+|v|^2= 1} . 
\ee
Every matrix  $g\in \SU{2}$ is determined by its first column, which is a vector of unit length in $\NC^2$. In fact, setting 
\begin{gather} 
\zx  =\qmat{u\\v}\in S^3\subset \NC^2, \quad \text{and}\quad  J \zx =\qmat{-\ol{v}\\\hp \ol{u}}, \\
\intertext{we may write} 
\label{eqn:SU(2)}
 g(\zx)=\qmat{ \zx & J \zx}=\qmat{u &-\ol{v}\\ v &\hp \ol{u}}. 
\end{gather}
Thus  $\SU{2}$ can be identified  with the unit sphere $S^3$ contained in $\NR^4\simeq \NC^2$. The polar coordinates in $\NR^4$, $\NR_+\times S^3 \ni (t,\zx) \mapsto t\zx \in \NR^4\setminus \{0\}$ become a bijection of $\NR^4\setminus \{0\}$ with $\NR_+\times \SU{2}$, the multiplicative group of nonzero quaternions.    

We take the customary Euler angles $(\zy,\,\zvf,\, \zc)$ as parameters for $\SU{2}$ and set  
$u=\cos(\zy/2)e^{i(\zvf+\zc)}$, $v=i\sin(\zy/2)e^{i(\zc-\zvf)}$, where $0<\zy<\zp$, $0\leq \zvf <2\zp$, $-2\zp\leq \zc <2\zp$,  
\begin{gather}\label{eqn:Euler_angles}
g= g(\zy,\,\zvf,\, \zc) = \qmat{\cos(\zy/2)e^{i(\zvf+\zc)/2} & i\sin(\zy/2)e^{i(\zvf-\zc)/2} \\ i\sin(\zy/2)e^{i(\zc-\zvf)/2} &\cos(\zy/2)e^{-i(\zvf+\zc)/2}}\nonumber \\[6pt]
= \qmat{e^{i\zvf/2 }& 0 \\0 & e^{-i\zvf/2 } } \qmat{\cos(\zy/2) &  i\sin(\zy/2) \\   i\sin(\zy/2)  & \cos(\zy/2) }  \qmat{e^{i\zc/2 }& 0 \\0 & e^{-i\zc/2 } } .
\end{gather}
In particular, the subgroup $K\subset \SU{2}$ of diagonal matrices of the form  
\be\label{eqn:diagonal-subgroup} 
d(\zvf)=g(0,\,2\zvf,\,0) =  \qmat{e^{i\zvf } & 0\\ 0 &e^{-i\zvf }},  \qquad  0\leq \zvf <2\zp, 
\ee
is isomorphic to $\gr{U}{1}$ and can be identified with the unit circle $S^1$ in the plane $\NC\simeq \NR^2$.

Let us recall the well-known identification  of $\NR^3$ with the space of  traceless Hermitian $2$ by $2$ matrices obtained  in terms of the Pauli matrices $\zs_\za$, 
%which to a vector $x=(x_1,\,x_2,\,x_3)\in \NR^3$ assigns a  Hermitian matrix $x\cdot \zs$ defined as 
\be\label{eqn:Pauli_mat} 
(x_1,\,x_2,\,x_3)  \longleftrightarrow x\cdot \zs=\IS{\za=1}{3}x_\za\zs_\za  %+x_2\zs_2+x_3\zs_3
= \qmat{x_3&x_1-ix_2\\x_1+ix_2&-x_3}.
%\zs_1=\pmat{0&1\\1&0},\quad \zs_2=\qmat{0&-i\\i&\hp 0},\quad \zs_3=\qmat{1&\hp 0\\0&-1}
\ee
Thus there is a well-defined action of  the group $\SU{2}$  on $\NR^3$ by means of 
%As Pauli matrices we take
%\be\label{eqn:Pauli_mat} 
%\zs_1=\pmat{0&1\\1&0},\quad \zs_2=\qmat{0&-i\\i&\hp 0},\quad \zs_3=\qmat{1&\hp 0\\0&-1}
%\ee
\begin{gather}
%x\cdot \zs=x_1\zs_1+x_2\zs_2+x_3\zs_3= \qmat{x_3&x_1-ix_2\\x_1+ix_2&-x_3}.  
%\intertext{and define its transform $g\cdot x$ under $g\in \SU{2}$ by the formula}
(g\cdot x)\cdot \zs=g^*(x\cdot \zs)g. \label{eqn:SU(2)-action}
\end{gather}
This is an orthogonal action, since $|x|=-\det x\cdot \zs = - \det g^*(x\cdot \zs)g = |g\cdot x|$, and gives the familiar representation of $\SU{2}$ by rotations of $\NR^3$. 
 
In particular, the orbit of $e_3=(0,\,0,\,1)\in \NR^3$ provides a convenient  identification of the unit sphere $S^2\subset\NR^3$ with the homogeneous (coset) space $\SU{2}/\gr{U}{1}$.
Thus for any $\zx \in S^3$ we may define an element  $H(\zx)\in \NR^3$  by the formula 
\begin{gather}
%g(\zx)^* \zs_3 g(\zx) =
 g(\zx)^*( e_3\cdot \zs)g(\zx) = H(\zx)\cdot \zs,  \quad \intertext{and since $|H(\zx)| = 1$,\ we actually   have a map}
 H : \zx \in S^3 \mapsto  H(\zx)\in S^2. \label{eqn:hopf_fib} 
\end{gather}
Moreover, since $ H (\zl\zx)=H(\zx)$ if  $|\zl|=1$, 
each orbit of $S^1$ in $S^3$ is mapped to a single point, hence $H$ is a fibration  of $S^3$ by means of $S^1$, called \emph{the Hopf fibration}. 

Let us now look at the complex projective space  $\NC{P^1}$ of (complex) dimension~$1$. 
Given $(\zx_0,\,\zx_1)\neq 0$ we denote by $l(\zx_0,\,\zx_1)$ the complex line in $\NC^{2}$ passing through $(\zx_0,\,\zx_1)$ and the origin $0$, and by $\zP$ the map (canonical projection) assigning the line $l(\zx_0,\,\zx_1)$ to that point $(\zx_0,\,\zx_1)$. Any pair of complex numbers $(\zx_0,\,\zx_1)$, where $\zx_0$ and 
$\zx_1$ are  not both equal to zero determining the line $l=l(\zx_0,\,\zx_1)$, is called a set of homogeneous coordinates of $l$ and denoted by $[\zx_0 \,:\,\zx_1]$. Thus the quotient $ \NC^{2}_*/\NC_*$, 
where the asterisk $*$ at the subscript position signifies the removal of $0$, may be identified with the set of one-dimensional complex subspaces in $\NC^2$ --- the complex projective space  $\NC{P^1}$. Furthermore, by restricting the projection $\zP$ to the unit sphere $S^3\subset \NC^{2}_*$, 
we see that $\zP:S^3\to \NC{P^1}$ is surjective, so we may identify $\NC{P^1}$ with $S^3/S^1$. 

On the other hand, it is customary to identify $\NC{P^1}$ with the compactified complex plane $\overline{\NC}=\NC\cup \{\infty\}$ by using the so-called inhomogeneous coordinates in  $\NC{P^1}$  
 \[
[\zx_0 \,:\,\zx_1]\mapsto \begin{cases}
                        	w=\dfrac{\zx_1}{\zx_0}, & \text{if}\ \zx_0\neq0\\
 				w'=\dfrac{\zx_0}{\zx_1}, & \text{if}\ \zx_1\neq0. 
                          \end{cases}
\]

We shall denote  by $\zq$ the map \fun{\zq}{\NC{P^1}}{\ol{\NC}} (identification of $\NC{P^1}$ with $\ol{\NC}=\NC\cup \{\infty\}$) given by the top line above, so that
\[
\zq([\zx_0 \,:\,\zx_1]) =  \begin{cases}
                        	 \dfrac{\zx_1}{\zx_0}, & \text{if}\ \zx_0\neq0\\
                        	 \infty, & \text{if}\ \zx_0=0.
                        	 \end{cases} 
\] 
From this  formula we get the familiar expression for the stereographic projection of $S^2$ from the South Pole $s\in S^2$ onto the complex plane
\[
w =\dfrac{\zx_1}{\zx_0}= \dfrac{x_1+ix_2}{1+x_3}. 
\] 
 The standard matrix action of the group $\gru{SL}{2}{C}$ on the space $\NC^{2}$ gives rise to the action on $\NC{P^1}$ 
\[ [\zx_0 \,:\,\zx_1]\mapsto g\cdot [\zx_0 \,:\,\zx_1]= [a\zx_0 +b \zx_1 \,:\,c\zx_0+d\zx_1] = [1\,:\,\frac{c\zx_0+d\zx_1}{a\zx_0 +b \zx_1}] ,
\]
inducing in turn the action on the compactified plane $\ol{\NC}$ by means of the homographic (linear fractional) maps 
\begin{equation}\label{basic_homography}
\zz\mapsto g\cdot \zz=\frac{c +d \zz }{a  +b  \zz },\qquad  \text{for}\quad g=\qmat{a&b\\c&d} \quad\text{and}\qquad a  +b \zz \neq 0 .  
\ee
This situation is summarized in the well-known commutative diagram
\begin{theo}\label{thm:Central_SU(2)}
Let   $H : S^3 \to S^2$ be the Hopf fibration, $\zP :S^3 \to  \NC{P^1}\simeq S^3/S^1 $  be the  projection of the sphere  $S^3$ on the projective space  corresponding to the orbit map and $\zF: \ol{\NC} \to S^2$ be the inverse mapping to the stereographic projection from the South Pole $s=-e_3=(0,\,0,\,-1)\in S^2$. 
The diagram 
\begin{gather}\label{eqn:Hopf_diagr}
\xymatrix{
{S^3 } \ar@{->}[r]^{H} \ar@{->}[d]_{\zP}    & {S^2 }  \\ 
{\NC{P^1}} \ar@{->}[r]^{\zq }&   \ol{\NC}  \ar@{->}[u]_{\zF} } 
\qquad \\
\textbf{\rm Diagram 1.} \nonumber
\end{gather}
is a commutative diagram of maps intertwining the respective actions of the group $\gr{SU}{2}$.
\end{theo}
 
A similar construction of the projective space $\cpn$ can be carried out at the general level (where $N$ is an arbitrary natural number). We define 
\begin{gather}
\NC{P^N}=\NC^{N+1}_*/\NC_*\simeq  \gr{SU}{N+1}/ \mathbf{S}(\gr{U}{1}\times \gr{U}{N}).
\end{gather}
The unit sphere $S^{2N+1}\subset \BC{N+1}$ also admits  the Hopf fibration $S^1\to S^{2N+1}\to \NC{P^N}$,  where the action of $S^1$ on $S^{2N+1}$ is the componentwise multiplication  by $\zl\in S^1$. 
The space $\NC{P^N}$ is equipped with a natural Riemannian metric, the so-called Fubini--Study metric, but we shall not need an explicit form of it. % --- see however \cite{PW-85}. 
The situation is to a great extent analogous to the previous one, and at the higher-dimensional level we also have the following diagram. 

\begin{Theo} \label{Theo2} 
Let $\zx$ be a homomorphism of $K\simeq  \gr{U}{1}$ into $\NC_*$, the multiplicative group of nonzero complex numbers, \ie\ a character of $\gr{U}{1}$. Then each smooth (i.e. infinitely differentiable) map \fun{F}{\SU{2}}{\NC^{N+1}} such that 
\be \label{eqn:covariant_map}
F(gh)=\zx(h)F(g),\qquad g\in \SU{2},\ h\in K
\ee
induces a unique smooth mapping  \fun{\zF}{S^2}{\NC{P^N}} by the relation $\zF(g\cdot e_1)=\zP(F(g))$. This can be expressed by means of the commutativity of the diagram 
\begin{gather} \label{diagram2}
\xymatrix{
{\SU{2}\simeq S^3}  \ar@{->}[r]^{F} \ar@{->}[d]_{H}   & {\NC^{N+1}}  \ar@{->}[d]_{\zP} \\ 
{S^2} \ar@{->}[r]^{\zF}&  {\NC{P^N}}   } 
\qquad \\
\textbf{\rm Diagram 2.} \nonumber
\end{gather}
\end{Theo}

\subsection{Coherent states and covariant maps}
Of fundamental importance for a quantum mechanical description of physical systems is that a state of a system is determined by a ray (one dimensional subspace) in a Hilbert space rather than a single vector. Hence the space of states is properly described as the projective space $PH$ of a certain Hilbert space $H$ rather than $H$ itself. 
In the case when the Hilbert space is finite-dimensional, by fixing a basis we may identify the projective space $PH$ with the standard projective space $\cpn$. 
 
The well-known Perelomov definition of generalized coherent states, cf. \cite{Pe},  is a natural source of  maps satisfying equation \eqref{eqn:covariant_map} (we shall refer to them as ``covariant maps'')  

\Bdef{Generalized coherent states (Perelomov)}Given a representation $T$ of a group $G$ in a Hilbert space $\CH$ and $\zc_0\in \CH$, such that $T(h)\zc_0= \za(h)\zc_0$, with $\za(h)\in S^1$ for $h$ belonging to a subgroup $H\subset G$, the image of the orbit $\set{T(g)\zc_0}{g\in G}$ in the set of states $\mathbf{P}(H)$ is said to be a system of (generalized) coherent states of type $(T,\,\zc_0)$.
\Edef 

In this paper we are going to describe an application of that construction in the context of $\cpn$ sigma models. 
 
\subsection{A brief review of irreducible representations of $\SU{2}$}

One realizes representations of the group $\gru{SL}{2}{\NC}$ on $\CP(\NC^2)$, the space of complex-valued polynomials in two complex variables $z_1,\,z_2$. The action is transferred from the  standard (right) matrix action of $\gru{SL}{2}{\NC}$ on  row vectors in $\NC^2$.  
For any polynomial $p(z)=p(z_1,z_2)$ in  $\CP(\NC^2)$ we set 
\be  \label{eqn:action on poly's}
g\cdot p(z) =  p(z  g) = p(az_1+cz_2,\,bz_1+dz_2),  
\quad  \text{for} \     g=\qmat{ a &b \\c & d}\ \in \gru{SL}{2}{\NC}.
\ee
 
Let  $d$ be a non-negative integer and $\CP^d=\CP^d(\NC^2)\subset \CP(\NC^2)$ be the subspace of homogeneous polynomials of degree $d$ in $z_1,\,z_2$. Clearly  $\dim \CP^d=d+1$. 
It is easy to observe that the action of $\gru{SL}{2}{\NC}$ leaves invariant each of the subspaces $\CP^d$ and hence the restriction of the action \eqref{eqn:action on poly's} to  $\CP^d$ defines a representation $\gru{SL}{2}{\NC}$  in  $\CP^d$. This action is known to be irreducible  for each integer $d$, and so is its restriction to the subgroup $\SU{2}\subset \gru{SL}{2}{\NC}$ of unitary matrices.  
Dealing with these representations we shall often employ the notation used in the quantum theory of spin, and parametrize representations by the value $j=d/2$ with the meaning of the total spin. Accordingly, the representation space  $\CP^{2j}$ will be denoted by $\CH^{(j)}$,  where $\dim \CH^{(j)}=2j+1$, and  for $g\in \SU{2}$  the  operators of the representation will be denoted $T^{(j)} (g)$.  Explicitly, with  $p\in\CH^{(j)}$ and  $z=(z_1,\,z_2)\in \NC^2$ we have
\be
T^{(j)}(g)p(z) = p(uz_1+vz_2,\,-\ol{v} z_1+\ol{u} z_2),\quad  \text{for}\     g=\qmat{u &-\ol{v}\\ v &\hp\ol{u}}\ \in  \SU{2}.   
\ee

\subsection{Matrix elements of irreducible representations of  $\SU{2}$}

It is easy to observe that polynomials of the form $z_1^{j+m}z_2^{j-m}$ are joint eigenvectors for the action of elements belonging to the diagonal subgroup $K$ in  $\SU{2}$, cf. \eqref{eqn:diagonal-subgroup}, with eigenvalue $\zq_m(d(\zvf))=e^{i2m\zvf}$. They are the weight vectors for the representation $T^{(j)}$. 
In the context of the quantum mechanical description of spin, the weight vectors  correspond to the states with fixed spin $j$ and projection $m$ of spin on the third axis, which are usually denoted by $|j,\, m\rangle$. 
 
We fix a basis of $\CH^{(j)}$ consisting of weight vectors normalized as follows:  
\be\label{eqn:weight_basis}
w^{(j)}_m(z)= \frac{z_1^{j+m}z_2^{j-m}}{[(j+m)!(j-m)!]^{1/2}},\qquad m=-j,\,-j+1,\,\ldots,\, j-1,\,j.  
\ee
This gives an orthonormal basis in  $\CH^{(j)}$ with respect to the inner product (Fischer inner product) which is defined in the following way. To every $p\in \CH^{(j)}$ we assign a differential polynomial (homogeneous partial differential operator) in the variables $z_1,\,z_2$, denoted $p(\p)$, by the following formula:  
\[
p(z)=\IS{\za+\zb=2j}{}p_{\za\,\zb}z_1^\za z_2^\zb, \quad \longrightarrow \quad 
p(\p)=\IS{\za+\zb=2j}{}p_{\za\,\zb}\p_1^\za\p_2^\zb.
\] 
Then, with the bar denoting complex conjugation, we set for $p,\,q \in \CH^{(j)}$
\be\label{Fischer_inn_prod}
(p\mid q) =\ol{q}(\p) p. 
  \ee
It can be verified  that this is indeed an inner product and that the set 
 $\{w^{(j)}_m\}$ is an orthonormal basis with respect to it. Moreover, the representation $T^{(j)}$ of the group $\SU{2}$ is unitary, cf. e.g.  \cite[Ch.~8]{Waw}.

The matrix elements $t^{(j)}_{km}(g)$ of the representation $T^{(j)}$   with respect to the basis  of the weight vectors $\{w^{(j)}_m \}$  are defined by the expansion
\be\label{eqn:matrix_elements}
T^{(j)}(g) w^{(j)}_m(z)=  \IS{k=-j}{j}t^{(j)}_{km}(g)w^{(j)}_k(z).
\ee
Using the parametrization \eqref{eqn:SU(2)} of $\SU{2}$ we have an explicit formula
\be\label{eqn:before_ext}
\begin{split}
t^{(j)}_{km}(g(\zx))=  [(j+m)!(j-m)!(j+k)!(j-k)!]^{-1/2}\times \\ 
\p_1^{j+k}\p_2^{j-k}(uz_1+vz_2)^{j+m}(-\ol{v} z_1+\ol{u} z_2)^{j-m} .  
\end{split}
\ee
This defines matrix elements $t^{(j)}_{km}$ as functions on the unit sphere $S^3$ in $\NR^4$, however one can extend them to homogeneous polynomials on $\NR^4$ by the formula
 \[
\widetilde{t}^{(j)}_{km}(x)= |x|^{2j}t^{(j)}_{km}(g(|x|^{-1}x)),\qquad \text{for}\ x\in \NR^4\setminus\{0\}.
\]
Remarkably enough, these extensions turn out to be harmonic polynomials. 
\begin{Theo}\label{thm:harm_exten}
The matrix elements $\zx\mapsto t^{(j)}_{km}(g(\zx))$ are spherical harmonics of degree $2j$, i.e. their homogeneous extensions to $\NR^4$ are harmonic polynomials.  
\end{Theo}
Before studying the matrix elements in their full generality we consider the special cases where $j=1$ or $j=3/2$.

\subsection{Spin $l=1$ case and classical Veronese surfaces}

We shall examine in this and the following section the coherent states systems associated with the weight vectors for the spin $1$ case and shall show that they are in fact the classical Veronese surface and its harmonic transforms.   
The physical content of this discussion will perhaps be clearer if we employ the notation of physical literature connected with the quantum mechanical description of spin. % for the case of spin $l=1$.
The weight vectors $w_{m}^{(j)}$ are usually denoted by $|j,\, m\rangle$, where $j$ is the total spin of the system, and $m$ is the spin projection. For $j=1$ we have  
\begin{align}
&|1, \,1\rangle \simeq w_{1}^{(1)}(z)=2^{-1/2}z_1^2;\qquad  \nonumber\\
&|1,\,0\rangle \simeq w_{0}^{(1)}(z)= z_1z_2 ;\qquad  \label{weights_for_l=1}\\
&|1,-1\rangle \simeq w_{-1}^{(1)}(z)= 2^{-1/2}z_2^2. \nonumber
\end{align}
Setting in  \eqref{eqn:matrix_elements} $j=1$ we obtain by a direct calculation the following form   
of the orbit maps $\SU{2}\ni g \mapsto T^{(1)}(g)|1,\, m\rangle \in \CH^{(1)} $ corresponding to the weight vectors  $|1,\, m\rangle $  %which give us basic instances of covariant maps fulfilling equation \eqref{eqn:covariant_map}. 
\begin{align} 
F_1(g (\zx))&=T^{(1)}(g(\zx)) |1, \,1\rangle =
 u^2|1, \,1\rangle +2^{1/2}{u}{v}|1, \, 0\rangle +{v^2}|1, -1\rangle ; \label{coeff_1} \\ 
F_0(g (\zx)) &=T^{(1)}(g(\zx)) |1, \,0\rangle =-2^{1/2}\ol{v}{u}|1, \,1\rangle  + 
 (|u|^2- |{v}|^2)|1, \,0\rangle \nonumber\\
&\hspace{3.1cm}+ 2^{1/2}{v}\ol{u}|1, -1\rangle ; \label{coeff_2}   \\
F_{-1}(g (\zx))&=T^{(1)}(g(\zx)) |1, -1\rangle = \ol{v}^2 |1, \,1\rangle  - 2^{1/2} {\ol{v}}{\ol{u}} |1, \, 0\rangle + \ol{u}^2 |1, -1\rangle . \label{coeff_3} 
\end{align}
Since  for  $d(\zvf)$ given by \eqref{eqn:diagonal-subgroup} and  $m=1,\,0,\,-1$ 
\[
F_{m}(g (\zx)d(\zvf))= e^{2im\zvf } F_{m}(g (\zx)),
\]
by identifying the space  $\CH^{(1)}$ with $\NC^3$ by means of the basis $\{|1,\, m\rangle\}$ and  referring to Diagram 2 of  Theorem 1 we see that each of these maps induces a map from the sphere $S^2$ to the projective space $\cp{2}$. Now setting $\zz=g\cdot 0= {v}/{u} \in \NC_*$ and  parametrizing the sphere by the complex plane we obtain the following result.    
\begin{theo} \label{eqns:Veron-seq}
The maps {\rm(\ref{coeff_1}--\ref{coeff_3})} in terms of coordinates  with respect to the basis $\{|1,\, m\rangle\}  $ coincide with what in  {\rm\cite{Boal}} is called the Veronese sequence  --- the \emph{Veronese surface} and its harmonic transforms:  
\begin{align}
\zz \mapsto \zf_1(\zz) &= [(1,\,2^{1/2} \zz,\, \zz^2)]\in \NC{P}^2; \label{align:coh_state1}\\
\zz \mapsto \zf_0(\zz) &= [(-2^{1/2}\ol{\zz},\,1- |\zz|^2,\, 2^{1/2}\zz)]\in \NC{P}^2;\label{align:coh_state0}\\
\zz \mapsto \zf_{-1}(\zz) &= [(\ol{\zz}^2,\,-2^{1/2} \ol{\zz},\, 1)]\in \NC{P}^2. \label{align:coh_state-1}
\end{align} 
\end{theo}
It may be worthwhile to express this map in terms of the Euler angles \eqref{eqn:Euler_angles}. Noting that 
\[
\zz = \frac{v}{u} = i \tan(\zy/2)e^{-i\zc},  % = \zz (\zy,\,\zc),
\]
it follows from $($\ref{align:coh_state1}--\ref{align:coh_state-1} $)$ that
\begin{align*}
\zf_1(\zy,\,\zc ) &= [(1,\,2^{1/2}i \tan(\zy/2)e^{-i\zc}, \, -2 \tan^2(\zy/2)e^{-i2\zc} )]\in \NC{P}^2; \\ %\label{align:coh_state1} 
\zf_0( \zy,\,\zc )  &=  [( 2^{1/2}i \tan(\zy/2)e^{i\zc}  ,\, 1-  \tan^2(\zy/2) ,  \, 2^{1/2}i \tan(\zy/2)e^{-i\zc} )]\in \NC{P}^2.   \\  
\zf_{-1}( \zy,\,\zc ) & =  [( -2 \tan^2(\zy/2)e^{i2\zc}  ,\,2^{1/2}i \tan(\zy/2)e^{i\zc}, \, 1)]\in \NC{P}^2.
 \end{align*} 
\begin{remark} 
The Veronese surface \eqref{align:coh_state1} is well known and much  studied in the  complex differential geometry. It  is a conformal minimal immersion  of $S^2$  into the projective space $\cp{2}$  with constant curvature $2$.     

The harmonic transforms  {\rm(\ref{align:coh_state0}--\ref{align:coh_state0})} of the Veronese surface  also represent conformal minimal immersions with constant curvature. However, unkike the first one, they are not holomorphic.

It is worth noting that the highest, lowest respective, weight vectors give rise to holomorphic, antiholomorphic respective maps into $\cp{2}$. 
 
The construction of  ``harmonic transforms'' was used in disguise and in a different context (sigma-models)  already in the classic paper of  Din and Zakrzewski  {\rm\cite{DZ}}.
\end{remark} 

In order to relate this with the formulation given in \cite{GG} we explicitly state the expressions for projector-valued functions corresponding to the coherent states.
\begin{Coro} For   $m=1,\,0,\,-1$  we denote 
\[
P_m(\zz)= T(g)|1,\, m\rangle \otimes  \langle m,\, 1| T^* (g), \qquad \text{where}\ \zz= g\cdot 0 .
\]
The projection-valued fields corresponding to the weight vectors $|1,\, m\rangle $ are as in \eqref{projectors1}
\begin{align*}
P_1(\zz) & = 
\dfrac{1}{(1+|\zz|^2)^2}
\qmat{1 & 2^{1/2}\ol{\zz} & \ol{\zz}^2\\
2^{1/2} \zz & 2 |\zz|^2 & 2^{1/2}\ol{\zz} |\zz|^2          \\
\zz^2 & 2^{1/2} \zz |\zz|^2 & |\zz|^4};
 \\[6pt] 
P_0(\zz) & = \dfrac{1}{(1+|\zz|^2)^2}
\qmat{ 2|\zz|^2 & -2^{1/2} \ol{\zz}(1- |\zz|^2) & -2 \ol{\zz}^2 \\
-2^{1/2} {\zz}(1- |\zz|^2)  & (1- |\zz|^2)^2 & 2^{1/2} \ol{\zz}(1- |\zz|^2) \\
-2 \zz^2 & 2^{1/2} \zz (1- |\zz|^2)  & 2|\zz|^2 };
\\[6pt]  
 P_{-1}(\zz) &  = 
\dfrac{1}{(1+|\zz|^2)^2}\qmat{
|\zz|^4 &-2^{1/2} \ol{\zz} |\zz|^2 & \ol{\zz}^2 \\
-2^{1/2} \zz |\zz|^2 & 2 |\zz|^2 & -2^{1/2} \ol{\zz} \\
\zz^2 &-2^{1/2} \zz &  1 }.
\end{align*}
\end{Coro}  
We note that our indexing here deviates from what was used in formulas following \eqref{projectors1} in Section $2$, since we want to conform with the customary ordering of weights. 
The present indices $m = 1,0,-1$ stand for $k = 0,1,2$ we used them in what follows. The projectors $P_m (\zz)$ coincide with the ones constructed from the recurrence relations \eqref{recurrence1} and \eqref{recurrence2}.
In fact, the operators $P_+$ and $P_{-}$ termed in \cite{GG} raising and lowering operators respectively, which are at the origin of  those recurrence relations can be modeled algebraically by means of the raising and lowering operators (shift operators) of the corresponding representation of $\SU2$.  

Recall that in the case of the spin $1$ representation, the shift operators act on the weight vectors \eqref{weights_for_l=1} as differential operators  
\[
\pi_{-}= 2^{-1/2}z_2\dfrac{d }{d{z_1}}; \qquad \pi_{+}= 2^{-1/2}z_1\dfrac{d }{d{z_2}}
\]
so that the weight vectors form a chain (a ladder) obtained by successive applications of $\pi_{+}$ or $\pi_{-}$
\begin{gather*}
\pi_{-}w_1= w_0,\qquad \pi_{-}w_0= w_{-1}\qquad  \pi_{-}w_{-1}= 0; \\
 \pi_{+}w_{-1}= w_0,\qquad \pi_{+}w_0= w_{1}\qquad  \pi_{+}w_{1}= 0. 
\end{gather*} 
Their matrices with respect to the basis of the weight vectors $(w_1,\, w_0,\,w_{1})$ are
\[
\Pi_{-}=\qmat{ 0 & 0 & 0 \\1& 0 &0 \\ 0& 1 & 0},\qquad 
\Pi_{+}=\qmat{ 0 & 1 & 0 \\0& 0 &1 \\ 0& 0 & 0}. 
\]
The covariant map with respect to the group action from one system to another may be obtained by sending $T(g)|1,\,-1\rangle $ to $T(g)|1,\,0\rangle $ and similarly $T(g)|1,\,0\rangle $ to $T(g)|1,\,1\rangle $. Formally we consider the maps $\Pi_{+}(g)= T(g)\Pi_{+} T(g)^*$ and  $\Pi_{-}(g)= T(g)\Pi_{-} T(g)^*$  and compute their matrices with respect to the basis of weight vectors. After somewhat tedious but straightforward computations we obtain
\[
\Pi_{+}(g)= \qmat{-2^{1/2}uv & \hp u^2 & 0\\-v^2 &\hp 0 & u^2\\\hp 0 & - v^2 & 2^{1/2}uv};\qquad
\Pi_{-}(g)= \qmat{-2^{1/2}\ol{uv} & -\ol{v}^2 & 0\\\ol{u}^2 &\ 0 & -\ol{v}^2\\  0 & \ol{u}^2 & 2^{1/2}\ol{uv}};
\] 
One can check that $\Pi_{+}(g): T(g)|1,\,0\rangle \mapsto T(g)|1,\,1\rangle $, or $\Pi_{-}(g) T(g)|1,\,1\rangle \mapsto T(g)|1,\,0\rangle $, explicitly
\begin{align*}
\Pi_{+}(g) T(g)|1,\,0\rangle  &= \Pi_{+}(g)\qmat{-2^{1/2}u\ol{v}\\ |u|^2-|v|^2\\ 2^{1/2}\ol{u}v} =\qmat{u^2\\ 2^{1/2}uv\\ v^2};\\
\Pi_{-}(g) T(g)|1,\,1\rangle  &= \Pi_{+}(g)\qmat{|u|^2\\2^{1/2}{u}v \\v^2}= 
\qmat{-2^{1/2}u\ol{v} \\ |u|^2-|v|^2 \\ 2^{1/2}\ol{u}v}. 
\end{align*}
To end this overview we restate, with the new indexing, the formulas \eqref{immersions1} for the  $\su{3}$-valued forms  $X_m$, which  give the immersions  into the Lie algebra $\su{3}$  of  the surfaces belonging to the Veronese sequence studied here and state their main geometrical properties.  
\begin{align}
\nonumber &X_1=\dfrac{i}{(1+\vert\zz\vert^2)^{2}}
\qmat{\frac{1}{3}(\vert\zz\vert^4+2\vert\zz\vert^2-2) & -2^{1/2}\ol{\zz} & -\ol{\zz}^2 \\
-2^{1/2}\zz & %\hspace{-5mm}
\frac{1}{3}(\vert\zz\vert^4-4\vert\zz\vert^2+1) & -2^{1/2}\vert\zz\vert^2\ol{\zz} \\
-\zz^2 & -2^{1/2}\vert\zz\vert^2\zz & %\hspace{-3mm}
-\frac{1}{3}(2\vert\zz\vert^4-2\vert\zz\vert^2-1)},\\
&X_0=\dfrac{i}{(1+\vert\zz\vert^2)^2}
\qmat{\vert\zz\vert^2-1 & -2^{1/2}\ol{\zz} & 0 \\
-2^{1/2}\zz & 0 & -2^{1/2}\ol{\zz} \label{immers-X}\\
0 & -2^{1/2}\zz & -\vert\zz\vert^2+1}, \\
\nonumber &X_{-1}=\dfrac{i}{(1+\vert\zz\vert^2)^2}
\qmat{-\frac{1}{3}(1-2\vert\zz\vert^4+2\vert\zz\vert^2) & -2^{1/2}\vert\zz\vert^2\ol{\zz} & \ol{\zz}^2 \\
-2^{1/2}\vert\zz\vert^2\zz & %\hspace{-7mm}
-\frac{1}{3}(1+\vert\zz\vert^4-4\vert\zz\vert^2) & -2^{1/2}\ol{\zz} \\
\zz^2 & -2^{1/2}\zz & %\hspace{-5mm}
-\frac{1}{3}(\vert\zz\vert^4+2\vert\zz\vert^2-2)}.
\end{align}

The angles between the  immersion functions $X_k$ and $X_l$ associated with the $\mathbb{C}P^2$ model are given by
\begin{table}[h]
\begin{tabular}{c|ccc}
$k\backslash l$ & 1 & 0 & -1\\
\hline 1 & $5/\sqrt{33}$ & $\sqrt{3/11}$ & 1/3 \\
0 & $\sqrt{3/11}$ & 9/11 & $\sqrt{3/11}$ \\
-1 & 1/3 & $\sqrt{3/11}$ & $5/\sqrt{33}$
\end{tabular}
\centering
\end{table}

We now explore certain geometrical characteristics of surfaces
\eqref{immers-X} immersed in the $\su{3}$ algebra   and express them in terms of the projectors $P_m$. 
Using the known expression for the Gaussian curvatures
\begin{equation}
 \mathcal{K}_m=-2\frac{\partial\ol{\partial}\ln\vert\tr(\partial P_m\cdot\ol{\partial}P_m)\vert}{\tr(\partial P_m\cdot\ol{\partial}P_m)},
\end{equation}
one checks easily that each of the surfaces $X_m$ has constant and positive curvature, that is
\begin{equation}
 \mathcal{K}_1=\mathcal{K}_{-1}=2,\qquad \mathcal{K}_0=1.
\end{equation}
Also the norm $\Vert\cdot\Vert=(\cdot,\cdot)^{1/2}$ of the mean curvature vectors
\begin{equation}
 \mathcal{H}_m=-4i\frac{[\partial P_m,\ol{\partial}P_m]}{\tr(\partial P_m\cdot\ol{\partial}P_m)}
\end{equation}
are constant and positive
\begin{equation}
 \Vert\mathcal{H}_1\Vert=\Vert\mathcal{H}_{-1}\Vert=4,\qquad \Vert\mathcal{H}_0\Vert=2.
\end{equation}
The Willmore functionals are defined by
\begin{equation}
 W_m=\int_{S^2}\tr([\partial P_m,\ol{\partial} P_m])^2d\zz^1d\zz^2,
\end{equation}
and by computing the integrals we get
\begin{equation}
 W_1=W_{-1}=4\pi,\qquad W_0=2\pi.
\end{equation}
The topological charges associated with the surfaces $X_m$ are defined by
\begin{equation*}
 Q_m=\frac{-2}{\pi}\int_{S^2}\tr(P_m\cdot [\partial P_m,\ol{\partial} P_m])d\zz^1d\zz^2, 
\end{equation*}
and are
\begin{equation*}
Q_1=2,\ \ Q_0=1,\ \ Q_{-1}=-2.
\end{equation*}
%that is
%\begin{equation}
 
%\end{equation}
The Euler--Poincar\'e characters are determined by
\begin{equation}
 \Delta_m=\frac{-2}{\pi}\int_{S^2}\partial\ol{\partial}\ln\vert\tr(\partial P_m\cdot\ol{\partial}P_m)\vert d\zz^1d\zz^2,
\end{equation}
and we obtain the same value for the surfaces $X_m$, i.e.
\begin{equation}
\Delta_1=\Delta_0=\Delta_{-1}=2.
\end{equation}
This means that the surfaces $X_m$ are homeomorphic to ovaloids, since $\mathcal{K}_m>0$.
\section{The case of spin $l=3/2$}
The  case of spin $l=3/2$ can be discussed in an analogous way as the previous one. The weight vectors $|{3}/{2},m\rangle$  may be identified with the polynomials 
\begin{equation*}
w^{(3/2)}_m(z)= \frac{z_1^{3/2+m}z_2^{3/2-m}}{[(3/2+m)!(3/2-m)!]^{1/2}},
\end{equation*}
hence by virtue of the formula \eqref{eqn:matrix_elements} the orbit of the highest weight vector $| {3}/{2},3/2\rangle$ can be parametrized as
\[
\begin{split}
S^3\ni %\zx=
\qmat{u\\v}\mapsto \sqrt{3!}u^3
\bigl(
\bigl| \tfrac{3}{2},\tfrac{3}{2}\bigr\rangle + \sqrt{3} \zz\bigl|\tfrac{3}{2}, \tfrac{1}{2}\bigr\rangle + \sqrt{3} \zz^2\bigl|\tfrac{3}{2}, -\tfrac{1}{2}\bigr\rangle \bigr. \\{} + \bigl. \zz^3\bigl|\tfrac{3}{2}, -\tfrac{3}{2}\bigr\rangle  \bigr), 
\end{split}
\]
where we have set as before $\zz= v/u$ , the stereographic parameter of the $S^2$ sphere. Thus the basic holomorphic solution of the $\cp{3}$ sigma model is obtained by projecting this orbit to the $\cp{3}$ space. This gives the Veronese immersion 
\[
\NC\ni \zz \mapsto (1, \sqrt{3} \zz , \sqrt{3}  \zz^2, \zz^3) \in \NC^4
\]
and  its corresponding  projector-valued field 
\arraycolsep 2pt
\begin{gather*} 
 P_{3/2}  =  \frac1  { (1 +  | \zz  | ^2 )^3}
	\qmat{
 1 & \sqrt{3} \ol{\zz } & \sqrt{3} \ol{\zz }^2 & \ol{\zz }^3 \\
 \sqrt{3} \zz  & 3 \left| \zz \right| ^2 & 3 \left| \zz \right| ^2 \ol{\zz } & \sqrt{3} \left| \zz \right| ^2 \ol{\zz }^2 \\
 \sqrt{3} \zz ^2 & 3 \zz  \left| \zz \right| ^2 & 3 \left| \zz \right| ^4 & \sqrt{3} \left| \zz \right| ^4 \ol{\zz } \\
 \zz ^3 & \sqrt{3} \zz ^2 \left| \zz \right| ^2 & \sqrt{3} \zz  \left| \zz \right| ^4 & \left| \zz \right| ^6 }.
\end{gather*} 
\arraycolsep 5pt
The corresponding $\su{4}$-valued immersion function $X_{3/2}$ takes the form
\arraycolsep 3pt%
\def\lleft{}
\def\rright{}
\begin{align*} 
X_{3/2} & =  %\\ 
 \frac{i}{(1 +  |\zz| ^2 )^3 } \left[
\begin{array}{cc}
  \frac{1}{4} (| \zz | ^6+3 | \zz | ^4+3 | \zz | ^2-3) & -\sqrt{3} \ol{\zz }\\  
 -\sqrt{3} \zz  & \frac{1}{4} (| \zz | ^6+3 | \zz | ^4-9 | \zz | ^2+1)   \\        
-\sqrt{3} \zz ^2 & -3 \zz  | \zz | ^2 \\
 -\zz ^3 & -\sqrt{3} \zz ^2 | \zz | ^2
\end{array} \right. \\
&\hspace{3.5cm}\left.\begin{array}{cc}
 -\sqrt{3} \ol{\zz }^2 & -\ol{\zz }^3 \\
-3 | \zz | ^2 \ol{\zz } & -\sqrt{3} | \zz | ^2 \ol{\zz }^2 \\
 \frac{1}{4} (| \zz | ^6-9 | \zz | ^4+3 | \zz | ^2+1) & -\sqrt{3} | \zz | ^4 \ol{\zz } \\
 -\sqrt{3} \zz  | \zz | ^4 & \frac{1}{4} (-3 | \zz | ^6+3 | \zz | ^4+3 | \zz | ^2+1) 
\end{array} \right] 
\end{align*}
\arraycolsep 5pt%
In an entirely analogous way one can compute the coherent state generated by the weight vector  $|{3}/{2},{1}/{2}\rangle$. It is parametrized by the following formula 
\[
\begin{split}
S^3\ni %\zx=
\qmat{u\\v}\mapsto
T(g)\bigl| \tfrac{3}{2},\tfrac{1}{2}\bigr\rangle 
= {}-
\sqrt{3!}u^2\ol{v}
\bigl| \tfrac{3}{2},\tfrac{3}{2}\bigr\rangle 
+ u(|u|^2-2|v|^2)
\bigl|\tfrac{3}{2}, \tfrac{1}{2}\bigr\rangle + {}\\
 v(2|u|^2-|v|^2) \bigl|\tfrac{3}{2}, -\tfrac{1}{2}\bigr\rangle \bigr. 
+  \sqrt{3!} v^2\ol{u}\bigl|\tfrac{3}{2}, -\tfrac{3}{2}\bigr\rangle  \bigr), 
\end{split}
\]
 After proper normalization we get
 \[
\NC\ni \zz \mapsto (3 |\zz|^2 , \sqrt{3} \zz (2|\zz|^2-1), \sqrt{3} \zz^2(|\zz|^2-2) , {}-3\zz^3) \in \NC^4
\]
and the projector-valued  field   corresponding to this imbedding is given by 
\arraycolsep 2pt%
\def\lleft{}
\def\rright{}
\begin{gather*} 
(1 +  |\zz| ^2 )^3 P_{1/2} ={} \\
 \qmat{ 
 3 \lleft| \zz \rright| ^2 & \sqrt{3} \lleft(2 \lleft| \zz \rright| ^2-1\rright) \ol{\zz } & \sqrt{3} \lleft(\lleft| \zz \rright| ^2-2\rright) \ol{\zz }^2 & -3 \ol{\zz }^3 \\
 \sqrt{3} \zz  \lleft(2 \lleft| \zz \rright| ^2-1\rright) & \lleft(1-2 \lleft| \zz \rright| ^2\rright)^2 & \lleft(\lleft| \zz \rright| ^2-2\rright) \lleft(2 \lleft| \zz \rright| ^2-1\rright) \ol{\zz } & -\sqrt{3} \lleft(2 \lleft| \zz \rright| ^2-1\rright) \ol{\zz }^2 \\
 \sqrt{3} \zz ^2 \lleft(\lleft| \zz \rright| ^2-2\rright) & \zz  \lleft(\lleft| \zz \rright| ^2-2\rright) \lleft(2 \lleft| \zz \rright| ^2-1\rright) & \lleft| \zz \rright| ^2 \lleft(\lleft| \zz \rright| ^2-2\rright)^2 & -\sqrt{3} \lleft| \zz \rright| ^2 \lleft(\lleft| \zz \rright| ^2-2\rright) \ol{\zz } \\
 -3 \zz ^3 & -\sqrt{3} \zz ^2 \lleft(2 \lleft| \zz \rright| ^2-1\rright) & -\sqrt{3} \zz  \lleft| \zz \rright| ^2 \lleft(\lleft| \zz \rright| ^2-2\rright) & 3 \lleft| \zz \rright| ^4  } 
\end{gather*}
The corresponding immersion function $X_{1/2}$ is obtained from \eqref{surfacesk} and reads
\begin{align*}
\hspace{-3cm}&X_{1/2} =  \frac{i}{\lleft(1 + \lleft| \zz \rright| ^2 \rright)^3} \left[
\begin{array}{cc}
 \frac{1}{4} \lleft(3 \lleft| \zz \rright| ^6+9 \lleft| \zz \rright| ^4-3 \lleft| \zz \rright| ^2-5\rright) & -\sqrt{3} \lleft(2 \lleft| \zz \rright| ^2+1\rright)\ol{\zz} \\
 -\sqrt{3} \zz  \lleft(2 \lleft| \zz \rright| ^2+1\rright) & \frac{1}{4} \lleft(\lleft(3 \lleft| \zz \rright| ^2-7\rright) \lleft| \zz \rright| ^4+\lleft| \zz \rright| ^2-1\rright) \\
 -\sqrt{3} \zz ^2 \lleft| \zz \rright| ^2 & -\zz  \lleft(2 \lleft| \zz \rright| ^4+\lleft| \zz \rright| ^2+2\rright) \\ 
 \zz ^3 & -\sqrt{3} \zz ^2 
\end{array}\right.\\
&\hspace{3.5cm}\left.\begin{array}{cc}
-\sqrt{3} \lleft| \zz \rright| ^2 \ol{\zz }^2 & \ol{\zz }^3 \\
\lleft(-2 \lleft| \zz \rright| ^4-\lleft| \zz \rright| ^2-2\rright) \ol{\zz } & -\sqrt{3} \ol{\zz }^2 \\
\frac{1}{4} \lleft(\lleft(1-\lleft| \zz \rright| ^2\rright) \lleft| \zz \rright| ^4-7 \lleft| \zz \rright| ^2+3\rright) & -\sqrt{3} \lleft| \zz \rright| ^2 \lleft(\lleft| \zz \rright| ^2+2\rright) \ol{\zz } \\
-\sqrt{3} \zz  \lleft| \zz \rright| ^2 \lleft(\lleft| \zz \rright| ^2+2\rright) & \frac{1}{4} \lleft(-5 \lleft| \zz \rright| ^6-3 \lleft| \zz \rright| ^4+9 \lleft| \zz \rright| ^2+3\rright) 
\end{array}
\right]\\
\end{align*}
\arraycolsep 5pt%
The remaining projection-valued fields and immersion functions are computed in an analogous way --- we just give results. 
\begin{align*}
&P_{-1/2} \ =  \frac{1}{\lleft(1 + \lleft| \zz \rright| ^2 \rright)^3} \left[
\begin{array}{cc}
 3 \lleft| \zz \rright| ^4 & \sqrt{3} \lleft| \zz \rright| ^2 \lleft(\lleft| \zz \rright| ^2-2\rright) \ol{\zz } \\
 \sqrt{3} \zz  \lleft| \zz \rright| ^2 \lleft(\lleft| \zz \rright| ^2-2\rright) & \lleft| \zz \rright| ^2 \lleft(\lleft| \zz \rright| ^2-2\rright)^2 \\
 -\sqrt{3} \zz ^2 \lleft(2 \lleft| \zz \rright| ^2-1\rright) & -\zz  \lleft(\lleft| \zz \rright| ^2-2\rright) \lleft(2 \lleft| \zz \rright| ^2-1\rright) \\
 3 \zz ^3 & \sqrt{3} \zz ^2 \lleft(\lleft| \zz \rright| ^2-2\rright) 
\end{array}\right.\\
&\hspace{4cm}\left.\begin{array}{cc}
-\sqrt{3} \lleft(2 \lleft| \zz \rright| ^2-1\rright) \ol{\zz }^2 & 3 \ol{\zz }^3 \\
\lleft(2-\lleft| \zz \rright| ^2\rright) \lleft(2 \lleft| \zz \rright| ^2-1\rright) \ol{\zz } & \sqrt{3} \lleft(\lleft| \zz \rright| ^2-2\rright) \ol{\zz }^2 \\
\lleft(1-2 \lleft| \zz \rright| ^2\rright)^2 & -\sqrt{3} \lleft(2 \lleft| \zz \rright| ^2-1\rright) \ol{\zz } \\
-\sqrt{3} \zz  \lleft(2 \lleft| \zz \rright| ^2-1\rright) & 3 \lleft| \zz \rright| ^2 
\end{array}
\right];\\[12pt]
\hspace{-3cm}&X_{-1/2} = \frac{i}{\lleft(1 + \lleft| \zz \rright| ^2 \rright)^3}\left[
\begin{array}{cc}
 \frac{1}{4} \lleft(5 \lleft| \zz \rright| ^6+3 \lleft| \zz \rright| ^4-9 \lleft| \zz \rright| ^2-3\rright) & -\sqrt{3} \lleft| \zz \rright| ^2 \lleft(\lleft| \zz \rright| ^2+2\rright) \ol{\zz } \\
 -\sqrt{3} \zz  \lleft| \zz \rright| ^2 \lleft(\lleft| \zz \rright| ^2+2\rright) & \frac{1}{4} \lleft(\lleft| \zz \rright| ^6-\lleft| \zz \rright| ^4+7 \lleft| \zz \rright| ^2-3\rright) \\
 \sqrt{3} \zz ^2 & -\zz  \lleft(2 \lleft| \zz \rright| ^4+\lleft| \zz \rright| ^2+2\rright) \\
 \zz ^3 & \sqrt{3} \zz ^2 \lleft| \zz \rright| ^2 
\end{array}\right.\\
&\hspace{3cm}\left.\begin{array}{cc}
\sqrt{3} \ol{\zz }^2 & \ol{\zz }^3 \\
-\lleft(2 \lleft| \zz \rright| ^4+\lleft| \zz \rright| ^2+2\rright) \ol{\zz } & \sqrt{3} \lleft| \zz \rright| ^2 \ol{\zz }^2 \\
\frac{1}{4} \lleft(-3 \lleft| \zz \rright| ^6+7 \lleft| \zz \rright| ^4-\lleft| \zz \rright| ^2+1\rright) & -\sqrt{3} \lleft(2 \lleft| \zz \rright| ^2+1\rright) \ol{\zz } \\
-\sqrt{3} \zz  \lleft(2 \lleft| \zz \rright| ^2+1\rright) & \frac{1}{4} \lleft(-3 \lleft| \zz \rright| ^6-9 \lleft| \zz \rright| ^4+3 \lleft| \zz \rright| ^2+5\rright) 
\end{array}
\right]
\end{align*}
and finally
\begin{align*}
&P_{-3/2} = \frac{1}{\lleft(1 + \lleft| \zz \rright| ^2 \rright)^3}
%\lleft( \begin{array}{cccc}
 \qmat{\lleft| \zz \rright| ^6 & -\sqrt{3} \lleft| \zz \rright| ^4 \ol{\zz } & \sqrt{3} \lleft| \zz \rright| ^2 \ol{\zz }^2 & -\ol{\zz }^3 \\
 -\sqrt{3} \zz  \lleft| \zz \rright| ^4 & 3 \lleft| \zz \rright| ^4 & -3 \lleft| \zz \rright| ^2 \ol{\zz } & \sqrt{3} \ol{\zz }^2 \\
 \sqrt{3} \zz ^2 \lleft| \zz \rright| ^2 & -3 \zz  \lleft| \zz \rright| ^2 & 3 \lleft| \zz \rright| ^2 & -\sqrt{3} \ol{\zz } \\
 -\zz ^3 & \sqrt{3} \zz ^2 & -\sqrt{3} \zz  & 1 }
\end{align*}  
\begin{align*}
\hspace{-3cm}&X_{-3/2} = \frac{i}{\lleft(1 + \lleft| \zz \rright| ^2 \rright)^3} \left[
\begin{array}{cc}
 \frac{1}{4} \lleft(3 \lleft| \zz \rright| ^6-3 \lleft| \zz \rright| ^4-3 \lleft| \zz \rright| ^2-1\rright) & -\sqrt{3} \lleft| \zz \rright| ^4 \ol{\zz } \\
 -\sqrt{3} \zz  \lleft| \zz \rright| ^4 & \frac{1}{4} \lleft(-\lleft| \zz \rright| ^6+9 \lleft| \zz \rright| ^4-3 \lleft| \zz \rright| ^2-1\rright) \\
 \sqrt{3} \zz ^2 \lleft| \zz \rright| ^2 & -3 \zz  \lleft| \zz \rright| ^2 \\
-\zz^3 & \sqrt{3}\zz^2
\end{array}\right.\\
&\hspace{3.5cm}\left.\begin{array}{cc}
\sqrt{3} \lleft| \zz \rright| ^2 \ol{\zz }^2 & -\ol{\zz }^3 \\
-3 \lleft| \zz \rright| ^2 \ol{\zz } & \sqrt{3} \ol{\zz }^2 \\
\frac{1}{4} \lleft(-\lleft| \zz \rright| ^6-3 \lleft| \zz \rright| ^4+9 \lleft| \zz \rright| ^2-1\rright) & -\sqrt{3} \ol{\zz } \\
-\sqrt{3} \zz  & \frac{1}{4} \lleft(-\lleft| \zz \rright| ^6-3 \lleft| \zz \rright| ^4-3 \lleft| \zz \rright| ^2+3\rright)
\end{array}
\right] 
\end{align*}
The angles between the immersion functions $X_k$ and $X_l$ associated with the $\mathbb{C}P^3$ model have the form

\begin{table}[h]
\begin{tabular}{c|cccc}
$k\backslash l$ & 3/2 & 1/2 & -1/2 & -3/2 \\
\hline 3/2 & 3/8 & 5/8 & 3/8 & 1/8 \\
1/2 & 5/8 & 11/8 & 9/8 & 3/8 \\
-1/2 & 3/8 & 9/8 & 11/8 & 5/8 \\
-3/2 & 1/8 & 3/8 & 5/8 & 3/8
\end{tabular}
\centering
\end{table}
\noindent The immersion functions $X_k$ are considered as position vectors whose endpoints trace out the two-dimensional surfaces in an $N^2-1$-dimensional $\su{N}$ algebra. This implies that the position vectors make a constant angle with each other, independent of the variables $\zz$ and $\ol{\zz}$. Furthermore, within a particular $\mathbb{C}P^{N-1}$ model and corresponding coherent state, the angle is the same for all choices of holomorphic solutions $P_k$ of the Euler-Lagrange equations (\ref{eq10}).

The Gaussian curvatures are positive and constant, that is
\begin{equation}
K_{3/2}=K_{-3/2}=\frac{4}{3},\qquad K_{1/2}=K_{-1/2}=4\frac{\sqrt{13}}{7}
\end{equation}
and the norm of the mean curvature vector are also positive and constant
\begin{equation}
\mathcal{H}_{3/2}=\mathcal{H}_{-3/2}=4,\qquad\mathcal{H}_{1/2}=\mathcal{H}_{-1/2}=4\frac{\sqrt{13}}{7}.
\end{equation}
The Willmore functionals are
\begin{equation}
 W_{3/2}=W_{-3/2}=\frac{9}{2}\pi,\qquad W_{1/2}=W_{-1/2}=\frac{13}{2}\pi.
\end{equation}
The topological charges take the form
\begin{equation}
Q_{3/2}=6,\qquad Q_{1/2}=2,\qquad Q_{-1/2}=-2,\qquad Q_{-3/2}=-6.
\end{equation}
The Euler-Poincar\'e characters are
\begin{equation}
\Delta_{3/2}=\Delta_{1/2}=\Delta_{-1/2}=\Delta_{-3/2}=4.
\end{equation}
This means that the surfaces $X_m$ associated with the $\mathbb{C}P^3$ model are homeomorphic to ovaloids in view that $K_m>0$.

\subsection{Explicit parametrization in terms of  Jacobi polynomials}
As is well known, c.f. \cite{Vil} or \cite{Waw} for example,, using the parametrization \eqref{eqn:Euler_angles} of  $\SU{2}$ by Euler angles one can express the matrix elements $t^{(l)}_{jk}$ in \eqref{eqn:before_ext} in terms of Jacobi polynomials $P^{(k-j, -j-k)}_{l+j}$. To be more precise, let us call the restrictions $ t^{(l)}_{jk}(g(\zy,\, 0,\, 0))$ of matrix elements to the subgroup  of  $\SU{2}$,  consisting of matrices $ \qmat{\cos(\zy/2) &  i\sin(\zy/2) \\   i\sin(\zy/2)  & \cos(\zy/2) }$, the reduced matrix elements. Then 
recalling that the vectors $\{w_m^{(l)}\}$  defined by \eqref{eqn:weight_basis} are the weight vectors with respect to the diagonal unitary group $K=\gr{U}{1}\subset \gr{SU}{2}$ and putting 
\begin{equation}\label{eqn:splitfactor}
\zt_{jk}(\zvf,\,\zc) = e^{2i(j\zvf+k\zc)}, \qquad \text{for}\quad  k,\,j = -l\,\ldots,\,l
\end{equation} 
we have  that 
\[
t^{(l)}_{jk}(g(\zy,\,\zvf,\,\zc)) = \zt_{jk}(\zvf,\,\zc) t^{(l)}_{jk}(g(\zy,\, 0,\, 0))
\]
and the reduced matrix elements  $t^{(l)}_{jk}(g(\zy,\, 0,\, 0))$ can be written  as  $P^{(k-j, -j-k)}_{l+j}(\cos\zy)$. 
For simplicity, we adopt here the  definition of the Jacobi polynomials 
$P^{(\za,\zb)}_k(x)$ with real parameters $(\za,\zb)$ by means of  the Rodrigues-type formula 
\be\label{Jacobi}
P^{(\za,\zb)}_k(x)= \frac{(-1)^k}{2^k\,k!} (1-x)^{-\za}(1+x)^{-\zb}\DER{k}{}{x}\Bigl[(1-x)^{k+\za}(1+x)^{k+\zb}\Bigr],
\ee
allowing  $\za,\zb$ to be arbitrary real parameters, since the usual assumption $\za,\,\zb>-1$ is needed to ensure integrability of the weight $(1-x)^{\za}(1+x)^{\zb}$ over the interval $(-1,\,1)$, which will not be the question here. Before discussing the general case, let us examine the cases of  spin $1$ and spin $3/2$ discussed above. 
\subsubsection{The case of spin $1$}
The Jacobi polynomials  $P^{(k-j,-k-j)}_{1+j}(x)$ for $j,\,k=-1,\,0,\,1$ are given by the following table: 
 \[ \begin{array}{c|c|c|c}
& k=1 & k=0 & k= - 1 \\[6pt] 
\hline 
j=\hp 1 & P^{(0, -2)}_2(x)= & P^{(-1, -1)}_2(x)= & P^{(-2, 0)}_2(x)= \\[6pt] 
 & \frac14(x+1)^2 &  \frac14(x^2-1) & \frac14(x-1)^2 \\
\hline  
j=\hp 0 & P^{(1, -1)}_1(x)=x+1  & P^{(0, 0)}_1(x)=x & P^{(-1, 1)}_1(x)= x-1  \\[6pt] 
\hline 
 j= -1	& P^{(2, 0)}_0(x)=1  & P^{(1, 1)}_0(x)=1  & P^{(0,2)}_0(x)=1 \\[6pt]
 %j= -1	& P^{(2, 0)}_0(x)=\frac14(1-x)^2 & P^{(1, 1)}_0(x)=\frac14(x^2-1) & P^{(0,2)}_0(x)=\frac14(1+x)^2. \\[6pt]
\hline  
\end{array}
\] 
It  is now straightforward to write down   expressions for the matrix elements involving the Jacobi polynomials.  % using the formulas in the quoted sources: 
\begin{alignat*}{2}
t^{(1)}_{11}(\zy,\,\zvf,\,\zc)      & = \cos^2(\zy/2) e^{i(\zc+\zvf)}   &   = & \hp \cos^{-2}(\zy/2)e^{i(\zc+\zvf)} P^{(0,-2)}_2(\cos\zy), \\
t^{(1)}_{01}(\zy,\,\zvf,\,\zc)     & = 2^{-1/2}i\sin \zy e^{i\zc}\   & = & \hp 2^{-1/2}i \dfrac{\sin(\zy/2)}{\cos(\zy/2)} e^{i\zc} P^{(1,-1)}_1(\cos\zy), \\
t^{(1)}_{{-1}1}(\zy,\,\zvf,\,\zc) & =  -  \sin^2(\zy/2) e^{i(\zc-\zvf)}\  & = & -\sin^2(\zy/2)e^{i(\zc-\zvf)} P^{(2,0)}_0(\cos\zy).  \\[12pt]
%\end{alignat*}
%\begin{alignat*}{2} %\label{second_column}
t^{(1)}_{10}( \zy,\,\zvf,\,\zc)    & = 2^{-1/2}i\sin \zy e^{i\zvf}  &   = & - 2^{3/2}i \sin ^{-1}\zy e^{i\zvf}  P^{(-1,-1)}_2(\cos\zy);  \\
t^{(1)}_{00}( \zy,\,\zvf,\,\zc)    & = \cos\zy &   = & \hp  P^{(0,0)}_1(\cos\zy);\\ 
t^{(1)}_{{-1}0}( \zy,\,\zvf,\,\zc) & = 2^{-1/2}i\sin \zy e^{-i\zvf} &  = &\hp  2^{-1/2}i\sin\zy e^{-i\zvf}  P^{(1,1)}_0(\cos\zy);\\[12pt]  %\label{third_column}
 t^{(1)}_{1-1}( \zy,\,\zvf,\,\zc)    & = - \sin^2(\zy/2)e^{i( \zvf-\zc)} &   = & \hp \sin^{-2}( \zy/2) e^{i( \zvf-\zc)}  P^{(-2,0)}_2(\cos\zy);  \\ 
t^{(1)}_{0-1}( \zy,\,\zvf,\,\zc)    & = 2^{-1/2}i\sin \zy e^{-i\zc}& = & -2^{-1/2}i \dfrac{\cos\zy/2}{\sin\zy/2} e^{-i\zc}  P^{(-1,1)}_1(\cos\zy); 
\\ t^{(1)}_{-1-1}( \zy,\,\zvf,\,\zc)    & = \cos^2(\zy/2)e^{-i(\zc+\zvf)}  &   = &  \hp \cos^2(\zy/2)e^{-i(\zc+\zvf)}   P^{(0,2)}_0(\cos\zy) .  
\end{alignat*}
 It may be  noted that the middle ($0$-th) column, i.e. $\{ t^{(1)}_{j0}( \zy,\,\zvf,\,\zc)  \}$   consists of standard spherical harmonics of degree $1$ (with respect to the variables $(\zc,\zvf)$).

\subsubsection{The case of spin $3/2$}{}

The Jacobi polynomials $P^{(k-j,-k-j)}_{\frac32+j}(x)$ relevant for this case are given in the following table. 
\[
\begin{array}{c|c|c|c|c}
%& k=2 \\ & k=3 & k=\frac12 & k= - \frac12 & k= - \frac32 \\[6pt]
& k=\frac32 & k=\frac12 & k= - \frac12 & k= - \frac32 \\[6pt]
\hline
j=\frac32 & P^{(0, -3)}_3(x) & P^{(-1, -2)}_3(x) & P^{(-2, -1)}_3(x) & P^{(-3, 0)}_3(x) \\[6pt]
& =\frac18(x+1)^3 & =\frac18(x-1)(x+1)^2 & =\frac18(1-x)^2(1+x) & =\frac18(x-1)^3 \\[6pt]
\hline
j=\frac12 & P^{(1, -2)}_2(x) & P^{(0, -1)}_2(x) & P^{(-1, 0)}_2(x) & P^{(-2, 1)}_2(x) \\[6pt]
& =\frac34(1+x)^2  & =\frac14(x+1)(3x-1) & =\frac14(1+3x)(x-1) & =\frac34(1-x)^2 \\[6pt]
\hline
j=-\frac12	& P^{(2, -1)}_1(x) & P^{(1, 0)}_1(x) & P^{(0,1)}_1(x) & P^{(-1,2)}_1(x)\\[6pt]
& =\frac32(1+x)  & =\frac12(3x+1)  & =\frac12(3x-1) & =\frac32(x-1) \\[6pt]
\hline
j=-\frac32 & P^{(3, 0)}_0(x)=1  & P^{(2, 1)}_0(x)=1 & P^{(1, 2)}_0(x)=1 & P^{(0, 3)}_0(x)=1 \\[6pt]
\hline
\end{array}
\]
  
In this case, for reasons of space, we give below only the reduced matrix elements, of which the full form can be obtained by combining the formulas for the  reduced  matrix elements $t^{(\frac32)}_{j k}(\theta,\, 0,\,0)$ with the factor $\zt_{jk}(\zvf,\zc)$ from \eqref{eqn:splitfactor}. 

\[
\begin{array}{ll}
t^{(\frac32)}_{\frac32 \frac32}(\theta) = \cos^3 \frac{\theta}{2}   &
t^{(\frac32)}_{\frac32 \frac12}(\theta) = \sqrt3 i \sin \frac{\theta}{2} \cos^2 \frac{\theta}{2}   \\
t^{(\frac32)}_{\frac12 \frac32}(\theta) = \sqrt3 i \sin \frac{\theta}{2} \cos^2 \frac{\theta}{2}  &
t^{(\frac32)}_{\frac12 \frac12}(\theta) = \cos^3 \frac{\theta}{2} - 2 \sin^2 \frac{\theta}{2} \cos \frac{\theta}{2}  \\
t^{(\frac32)}_{-\frac12 \frac32}(\theta) = -\sqrt3 \sin^2 \frac{\theta}{2} \cos \frac{\theta}{2}   &
t^{(\frac32)}_{-\frac12 \frac12}(\theta) = i (2 \sin \frac{\theta}{2} \cos^2 \frac{\theta}{2} - \sin^3 \frac{\theta}{2})  \\
t^{(\frac32)}_{-\frac32 \frac32}(\theta) = -i \sin^3 \frac{\theta}{2}   &
t^{(\frac32)}_{-\frac32 \frac12}(\theta) = -\sqrt3 \sin^2 \frac{\theta}{2} \cos \frac{\theta}{2}   \\
\\
t^{(\frac32)}_{\frac32 -\frac12}(\theta) = \sqrt3 i \sin \frac{\theta}{2} \cos^2 \frac{\theta}{2}  &
t^{(\frac32)}_{\frac32 -\frac32}(\theta) = -i \sin^3 \frac{\theta}{2}   \\
t^{(\frac32)}_{\frac12 -\frac12}(\theta) = \cos^3 \frac{\theta}{2} - 2 \sin^2 \frac{\theta}{2} \cos \frac{\theta}{2}    &
t^{(\frac32)}_{\frac12 -\frac32}(\theta) = -\sqrt3 \sin^2 \frac{\theta}{2} \cos \frac{\theta}{2}  \\
t^{(\frac32)}_{-\frac12 -\frac12}(\theta) = i (2 \sin \frac{\theta}{2} \cos^2 \frac{\theta}{2} - \sin^3 \frac{\theta}{2})   &
t^{(\frac32)}_{-\frac12 -\frac32}(\theta) = \sqrt3 i \sin \frac{\theta}{2} \cos^2 \frac{\theta}{2}   \\
t^{(\frac32)}_{-\frac32 -\frac12}(\theta) = -\sqrt3 \sin^2 \frac{\theta}{2} \cos \frac{\theta}{2}  &
t^{(\frac32)}_{-\frac32 -\frac32}(\theta) = \cos^3 \frac{\theta}{2}  
\end{array}
\]

\subsection{Matrix elements of $\SU{2}$ irreducible representations and general Veronese immersions} 

 We return to the general case of  the spin $j$ representation of $\SU{2}$, acting in the space  $\CH^{(j)}=\CP^{2j}(\NC^2)$ with $\dim \CH^{(j)}=2j+1$ which is  given by the formula
\[  %\label{eqn:SU(2)-action on poly's}
(T^{(j)}(g) p)(z_1,\,z_2) = p( uz_1+vz_2,\,-\ol{v}z_1+\ol{u}z_2), \quad \text{for}
\quad p \in \CH^{(j)}.
\]
The matrix elements %$t^{(j)}_{km}(g)$ of this representation are given by
\[
t^{(j)}_{km}(g)=(w_k^{(j)}\mid  T^{(j)}(g) w_m^{(j)})
\]
can be given in the following explicit form.
\begin{Theo}[cf. \cite{Vil,Waw}] Given any  half-integer $j$ and $-j\leq k,\, m\leq j$ set $\za= k-m,\ \zb= k+m$ and $n=j-k$. The matrix elements of the representation $T^{(j)}$ are given by
\begin{gather}\label{gen_elem} 
t^{(j)}_{km}(g)=t^{(j)}_{km}(g(\zy,\,\zvf,\,\zc)) = \zt_{km}(\zvf,\,\zc) t^{(j)}_{km}(g(\zy,0,0) )
\intertext{where}  
t^{(j)}_{km}(g(\zy,0,0) )= c(j,k,m)(\cos(\zy/2))^{-\zb}(\sin(\zy/2))^\za P_n^{(\za,-\zb)}(\cos\zy).
\end{gather}
and
\[
c(j,k,m)= i^{m-k}\frac{[(j+k)!(j-k)!]^{1/2}}{[(j+m)!(j-m)!]^{1/2}}.
\]
\end{Theo}
Since for any $m$ the coherent state map 
\[
\gr{SU}{2} \ni g \mapsto (t^{(j)}_{-j\,m}(g ),\,t^{(j)}_{-j+1\,m}(g ),\, \ldots,\, t^{(j)}_{j\,m}(g) )\in \NC^{2j+1} 
\]
satisfies the assumptions of Theorem \ref{Theo2}, due to Diagram 2, it induces an imbedding of the sphere $S^2$ into the projective space $\cp{2j}$. Comparing with the results of the paper \cite{Boal},  
we see that it gives a conformal minimal imbedding belonging to the Veronese family.  

It might be interesting to investigate the implications of the direct parametrization of this map on the study of the geometry in question. Some work in this direction is in progress.  

%\newpage
\section{Final remarks and future developments}

The links between different analytic descriptions of $SU(2)$ coherent states and the $\cpn$ sigma models (defined on the Riemann sphere with finite actions) can be generalized to more general sigma models than the one proposed in this paper. An analysis of the complex Grassmannian sigma models taking values on the homogeneous spaces
\begin{equation}
 G(m,n)=\frac{SU(N)}{S(U(m)\times U(n))},\qquad N=m+n
\end{equation}
similar to the one carried out in section 2 can provide us with a more general explicit form for coherent states. These models share many common properties with the $\cpn$ models presented here. Namely, they possess an infinite number of local and/or nonlocal conserved quantities, as well as infinite-dimensional dynamical symmetries generating the Kac--Moody algebra. Both the Grassmannian sigma model equations and the $\cpn$ sigma model have a Hamiltonian structure, complete integrability, and the existence of multisoliton solutions, where the linear spectral problem is well established \cite{WZ}. Several classes of solutions of both equations are known. These solutions can be expressed in terms of holomorphic functions and functions obtained from them by a procedure similar to the one presented in this paper, which allows us to generate a complete set of solutions (more general than the ones constructed from the $\cpn$ model). It is evident that our approach can be applied to the complex Grassmannian sigma model which can describe much more diverse types of coherent states. This task will be undertaken in our future work.

\section*{Acknowledgments}
AMG's work was supported partially by a research grant from NSERC of Canada and by the Unit\'e Mixte Internationale (UMI) du FQRNT. AS wishes to acknowledge and thank the Mathematical Physics Laboratory of the Centre de Recher\-ches Math\'ematiques for their hospitality during his visit to the Universit\'e de Montr\'eal. A preliminary version of this paper was presented by AS at the conference on Exact Solvability and Symmetry Avatars in Honour of Luc Vinet given at the Universit\'e de Montr\'eal.

\end{document}